\documentclass[sigconf,authorversion,nonacm]{acmart}

\usepackage{booktabs} % For formal tables

\setcopyright{none}
\usepackage{hyperref}
\def\etal{\textit{et al.}}

\usepackage{tabularx}
\begin{document}

% UPDATE the title, author names, affiliations and shortauthors 
\title{HDRVideo-GAN: Deep Generative HDR Video Reconstruction}
% \titlenote{Produces the permission block, and copyright information}

 \author{Mrinal Anand}
 \authornote{denotes equal contribution}
%   \orcid{1234-5678-9012}
  \affiliation{%
    \institution{Indian Institute of Technology}
    % \streetaddress{P.O. Box 1212}
    \city{Gandhinagar}
    \state{Gujarat}
    \country{India}
    % \postcode{43017-6221}
    % \postcode{43017-6221}
  }
  \email{mrinal.anand@iitgn.ac.in}
  
  \author{Nidhin Harilal}
  \authornotemark[1]
  \affiliation{%
    \institution{University of Colorado}
    % \streetaddress{P.O. Box 1212}
    \city{Boulder}
    \state{Colorado}
    \country{USA}
  }
  \email{nidhin.harilal@colorado.edu}
  
  \author{Chandan Kumar}
  \authornotemark[1]
  %\authornote{The secretary disavows any knowledge of this author's actions.}
  \affiliation{%
    \institution{Indian Institute of Technology}
    % \streetaddress{P.O. Box 1212}
    \city{Gandhinagar}
    \state{Gujarat}
    \country{India}
    % \postcode{43017-6221}
  }
  \email{chandan.kumar@alumni.iitgn.ac.in}
  
  \author{Shanmuganathan Raman}
  %\authornote{The secretary disavows any knowledge of this author's actions.}
  \affiliation{%
    \institution{Indian Institute of Technology}
    % \streetaddress{P.O. Box 1212}
    \city{Gandhinagar}
    \state{Gujarat}
    \country{India}
    % \postcode{43017-6221}
  }
  \email{shanmuga@iitgn.ac.in}

% The default list of authors is too long for headers.
\renewcommand{\shortauthors}{Anand \etal}

\begin{abstract}
High dynamic range (HDR) videos provide a more visually realistic experience than the standard low dynamic range (LDR) videos. Despite having significant progress in HDR imaging, it is still a challenging task to capture high-quality HDR video with a conventional off-the-shelf camera. Existing approaches rely entirely on using dense optical flow between the neighboring LDR sequences to reconstruct an HDR frame. However, they lead to inconsistencies in color and exposure over time when applied to alternating exposures with noisy frames. In this paper, we propose an end-to-end GAN-based framework for HDR video reconstruction from LDR sequences with alternating exposures. 
We first extract clean LDR frames from noisy LDR video with alternating exposures with a denoising network trained in a self-supervised setting. Using optical flow, we then align the neighboring alternating-exposure frames to a reference frame and then reconstruct high-quality HDR frames in a complete adversarial setting. To further improve the robustness and quality of generated frames, we incorporate temporal stability-based regularization term along with content and style-based losses in the cost function during the training procedure. Experimental results demonstrate that our framework achieves state-of-the-art performance and generates superior quality HDR frames of a video over the existing methods. 

\end{abstract}

% UPDATE THIS
% The code below should be generated by the tool at
% http://dl.acm.org/ccs.cfm
% Please copy and paste the code instead of the example below.
%



\maketitle

\section{Introduction}
\label{sec:introduction}
The dynamic range that the human visual system can experience in the real world is vast. Unfortunately, most off-the-shelf digital cameras capture only a limited range of illumination in the scene. This discrepancy has lead to a great deal of research in reconstructing still HDR images from the conventional LDR off-the-shelf camera images. Most of these works have used a bracketed exposure imaging method~\cite{mann1994beingundigital,debevec2008recovering,mann2002painting,kalantari2017deep} which involves taking multiple images at different exposures and merging them to generate a single HDR image.

Generating HDR images by taking multiple images with different exposures may involve object/ camera movement. Therefore these methods end up producing ghosting artifacts in dynamic scenes of the image. These artifacts can be reduced through various methods like replacing/rejecting the pixels that move across the images~\cite{khan2006ghost,zhang2010denoising,zheng2013hybrid}, merging all different exposure images with a reference image \cite{ward2003fast,tomaszewska2007image,gallo2015locally} or aligning and reconstructing in a unified optimization system \cite{sen2012robust,zheng2013hybrid}. Capturing HDR video directly involves expensive specialized cameras that use complex optical systems \cite{tocciversatile}, and sensors \cite{zhao2015unbounded}.

On the other hand, reconstructing HDR video from the LDR sequence obtained from a standard off-the-shelf camera is a much more challenging task. Existing methods that are focused on reconstructing HDR images have been observed to generate temporally unstable results when applied to video sequences.

There exist few works addressing this problem~\cite{kang2003high,mangiat2011spatially,kalantari2013patch,li2016maximum}, which are typically slow and have limitations in several scenarios. Recently, the first deep learning-based approach was proposed by Kalantari~\emph{et al.}~\cite{kalantari2019deep} for HDR video reconstruction, which utilized dense frame-to-frame motion information (optical flow)~\cite{original_flow}. Their method first aligns the neighboring alternating exposure LDR frames to a reference frame by computing the optical flow between them, and then they use a convolutional neural network (CNN) based model to merge and reconstruct the final HDR frame. Although they show a reduction in time while reconstructing the HDR frames by a certain factor but as pointed out by its authors, their approach still suffers from discoloration and flickering artifacts in the reconstructed HDR video frames. 

In this paper, we take inspiration from \cite{kalantari2019deep} and design a Generative Adversarial Network (GAN) based framework for reconstructing HDR video frames from the LDR sequence with alternating exposures. Kalantari~\emph{et al.}~\cite{kalantari2019deep} performed an end-to-end training by minimizing the error between the reconstructed and ground truth HDR video frames on a set of training scenes. We show that merely reducing the pixel to pixel error between reconstructed and ground truth HDR frames from noisy LDR is prone to content loss and undesirable artifacts in the generated frames. We address this by proposing a framework comprising of an LDR denoising network, a light-weight optical flow estimation network, and a GAN based model for final HDR reconstruction. We modify the training procedure of our GAN based model by incorporating a temporal-stability based regularization term~\cite{eilertsen_mantiuk_unger_2019} along with content and style-based losses in the cost function while training the network. We use an altered version of an existing optical flow estimation model, LiteFlowNet~\cite{hui18liteflownet}, which is fine-tuned to estimate the dense optical flow between LDR frames with varying exposures. The estimated optical flow is then used to align the LDR frames with alternating exposures to the current frame. The final HDR frame is generated by using a Generative Adversarial Network (GAN). In addition to the regularization term, standard adversarial loss, and the HDR reconstruction losses, we also incorporate perceptual loss~\cite{Johnson2016Perceptual}, and style-aware content loss~\cite{sanakoyeu2018styleaware} while training the network for better performance. The proposed framework generates temporally stable HDR video with high visual quality.

GAN based models require a lot of data for image synthesis tasks. For the HDR reconstruction, we generate our training dataset synthetically by extracting the input LDR frames from a set of open-sourced HDR video repositories~\cite{froehlich2014creating, kronander2014unified}. However, unlike these synthetically generated LDR videos, frames captured from standard digital cameras have varied noise in them. Therefore, for the framework to generalize well, we fuse the LDR frames with the Gaussian noise of varied signal-to-noise (SNR) ratios.  
The main contributions of our work are given below:
\begin{itemize}
\itemsep0em 
    \item We propose the first GAN-based method for the HDR video reconstruction by using LDR frames with alternating exposures. Our proposed framework consists of a denoising network for extracting clean LDR frames, a light-weight optical flow estimation network, and a GAN based model for final HDR reconstruction.  
    % from the noisy LDR frames with alternating exposures. 
    \item  We incorporate perceptual as well as style-aware content losses to improve the visual quality of HDR frames. Along with utilizing the optical flow, we also incorporate a temporal-stability based regularization while training to further reduce the temporal incoherence in the reconstructed HDR frames.
    \item Our experimental results on the different HDR video datasets demonstrate that the proposed framework outperforms the existing approaches and produces high-quality HDR video.
\end{itemize}

\noindent \textit{Outline of the paper:} The entire paper is organized as follows. Section \ref{related_work} narrates the related work in the area of HDR imaging. Section \ref{dataset_section} describes the dataset used for the experimentation. Section \ref{archi} presents the model architecture in detail. Section \ref{training_det} describes the details about the used hyperparameters. Section \ref{results} reports the qualitative and quantitative evaluations against the baseline. Finally, section \ref{sec:ablation_study} presents the ablation study.

\section{Related Work}
\label{related_work}

\begin{table}[!b]

\begin{tabularx}{\linewidth} {lX}

\toprule \toprule
Notation & Desciption\\
\midrule
\raggedright
% \multirow{12}{.3\linewidth}{AA and BBBB }& aaaaa bbbb ccccd ddeeeef fffggggh \\
$L_{i}$ -- & original $i^{th}$ LDR frame (alternating expos.)\\
$\hat{L}_{i}$ -- &  $i^{th}$ generated clean LDR (alternating expos.)\\
$\widetilde{L}_{i}$ -- &  aligned $i^{th}$ clean LDR (alternating expos.)\\
$H_{i}$ -- & Original $i^{th}$ HDR frame \\
$\widetilde{H}_{i}$ -- & Generated $i^{th}$ HDR frame \\
$T_{i}$ -- & $i^{th}$ Tonemapped Frame of original HDR\\
$\widetilde{T}_{i}$ -- & $i^{th}$ Tonemapped Frame of generated HDR\\

\bottomrule
\noalign{\vskip 0.1cm}
\end{tabularx}
 \caption{Description of notations frequently occurring in the paper}
\label{tab:notation}
\vspace{-0.5cm}
\end{table}

In the last few years with the onset on learning algorithm, the problem of HDR imaging has also been extensively explored. However, a lot of work is centered around the generation of still HDR images. One set of approaches uses a sequence of different exposure images to generate HDR images~\cite{debevec1997recovering, sen2012robust, hu2013hdr, oh2014robust, ma2017robust, kalantari2017deep}, while the other approach uses burst images to generate the HDR image~\cite{liu2014fast, hasinoff2016burst}. There are some more focused works in the last few years to generate HDR images from a single image~\cite{eilertsen2017hdr, marnerides2018expandnet}. Almost all these approaches are not suitable for generating HDR video because of the lack of temporal consistency in still HDR imaging. For brevity, we will only discuss the works related to the generation of HDR video.

The system that produced the most high-quality results to date has been the specialized camera that directly captures HDR video. These cameras include special sensors that can capture extensive dynamic range~\cite{brajovic1996sorting, seger1999hdrc, nayar2000high} or the camera which has beam-splitters that deflects the light to many sensors such that ever sensors measure the different amount of radiance concurrently. However, these approaches are limited because they need specialized custom hardware that has enormous costs and, therefore, less widespread~\cite{tocci2011versatile, kronander2013unified}.

One way to generate HDR video is from the input sequence of frames having alternate exposure of each frame. Kang~\emph{et al.}~\cite{kang2003high} first proposed the method of HDR video reconstruction using the alternating exposure LDR frames. They used optical flow to align the neighboring frames to the reference frame. After aligning the nearby frame to the reference frame, they take a weighted sum to combine with the reference frame to avoid ghosting artifacts. However, their approach leads to ghosting artifacts when the scene has a significant amount of motion.

Mangiant and Gibson~\cite{kang2003high} improved the approach of Kang~\emph{et al.}~\cite{kang2003high} using a block-based motion estimation method  which was coupled with a refinement stage. In their successive work, they filtered the region with a significant motion to minimize the blocking artifacts. However, their approach still had the blocking artifacts when the scene has substantial movement in it. In addition to that, their approach is limited to working only on two exposure sequences. Kalantari~\emph{et al.}~\cite{kalantari2013patch} propose a patch-based method to reconstruct the missing exposure at each frame. After reconstruction, all the images produced were combined to obtain the final HDR frame. Temporal coherency is improved by estimating the motion between the neighboring/adjacent and the reference frame. However, the patch search was constraint only to small window around the predicted movement, where a greedy approach obtains the window size. This method produces the result, which is significantly better than the previous approach. However, to solve the complex patch-based optimization was a time consuming process and produce a single HDR frame. A major drawback of this approach was that it often was unable to constrain the patch search properly and underestimates the search window size. Ghosting artifacts were observed in such cases.
One further work is \cite{gryaditskaya2015motion} that improves the method of Kalantari~\emph{et al.}~\cite{kalantari2013patch} by adaptively adjusting the exposure. In a recent work by Li~\emph{et al.}~\cite{li2016maximum} proposes to consider the HDR video reconstruction problem as a maximization of posteriori estimate. Their method focuses on finding the foreground and background of each HDR frame separately. They extract the background and foreground using rank minimization and multiscale adaptive regression techniques, respectively. The major drawback of this method is also the computational cost involved, it takes around two hours to generate a single frame with $1280\times720$ resolution. Additionally, many of the frames were accompanied by noise and discoloration. 

Recently, Kalantari and Ramamoorthi \cite{kalantari2019deep} have proposed an approach in which they have used two connected networks called Flow network and Merge Network. Flow Network aligns the neighboring frame with the current frame, while the Merge network is used to merge the aligned frames with the reference frame. Their approach solves the problem in the best way so far. This is also state of the art for HDR video generation. But there are still ghosting artifacts in the challenging cases where the reference image is overexposed or if there are notable parallax and occlusion.

\section{Dataset}
\label{dataset_section}

\begin{figure}[!b]
\centering
\includegraphics[width=1\columnwidth]{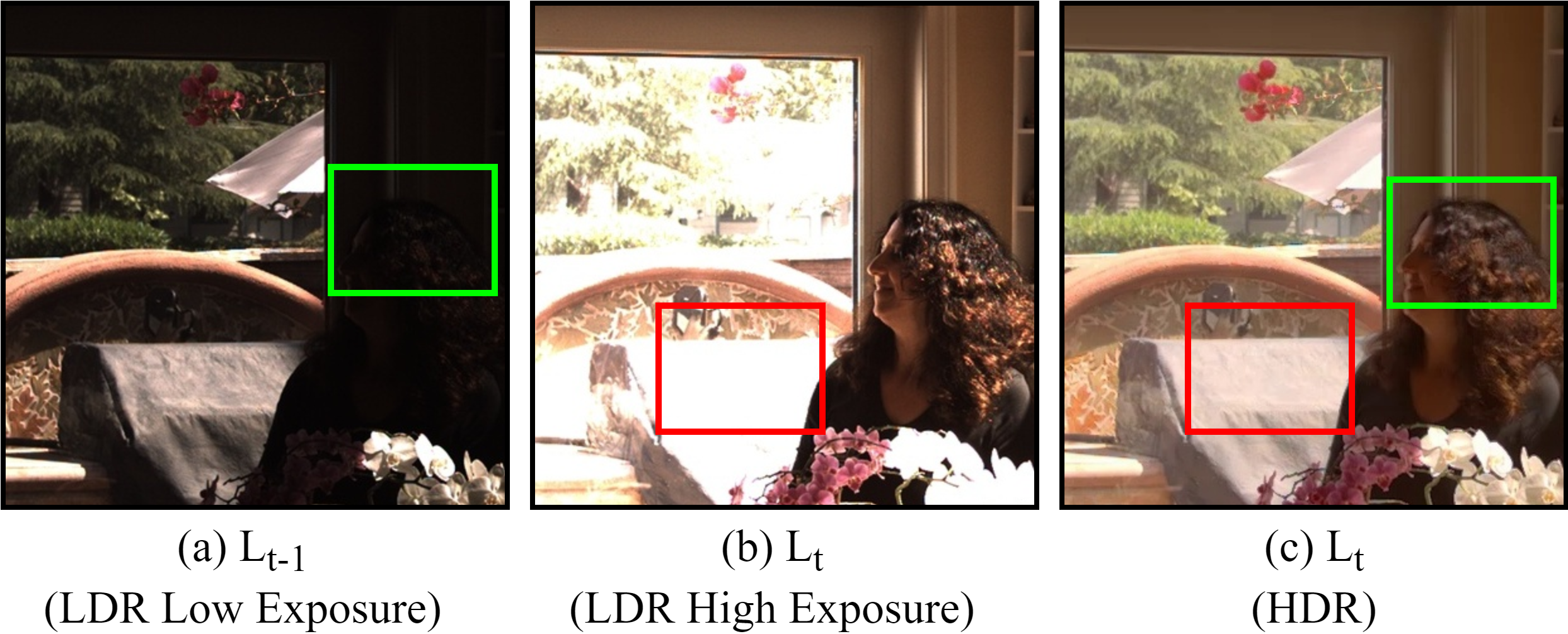}
\caption{Visual comparison of the over and under exposed LDR frames generated using Equation~\ref{eq1} with the corresponding HDR frame. Note the loss of details in the dark and bright regions of under and over exposed LDR frames.}
\label{expose}
\end{figure}

We require a large dataset consisting of HDR video frames with corresponding LDR frames having alternating exposures. We use two publicly available HDR video datasets curated by Froehlich~\emph{et al.}~\cite{froehlich2014creating} (13 videos) and Kronander~\emph{et al.}~\cite{kronander2014unified} (8 videos). These datasets were prepared using cameras with a specific optical design containing external~\cite{froehlich2014creating} and internal~\cite{kronander2014unified} beam filters. The dataset contains 41 HDR videos, out of which 38 were used for training, and the remaining three were used as the hold-out test set consistent with the current state-of-the-art method proposed by Kalantari~\emph{et al.}~\cite{kalantari2019deep} during experiments. We generate synthetic LDR frames from ground truth HDR frames at different exposures using Eq \ref{eq1}. 
\begin{equation} \label{eq1}
\begin{split}
L_{i}=g_i({H}_{i})=clip[({H}_{i}t_i)^{1/\gamma}]
\end{split}
\end{equation}
\noindent here $\gamma$ is 2.2, ${H_i}$ is HDR image in linear domain, $t_i$ is the exposure time, and $clip$ function clips the output in the range $[0,1]$. Figure~\ref{expose} shows the comparison of an underexposed, and overexposed LDR frames, along with an HDR frame. It can be clearly seen from Figure~\ref{expose} (a) and (b) clearly depicts the content loss in overexposed and underexposed LDR frames as compared to the corresponding HDR.

\section{Proposed Architecture}
\label{archi}

In this section, we present a detailed discussion of our proposed framework. Our proposed framework consists of three parts, i.e., an LDR denoising network for extracting clean LDR frames from noisy LDR video, a lightweight optical flow estimation network, and a GAN based model for the final reconstruction of high-quality HDR frames.

\textbf{Notations.} Table \ref{tab:notation} summarises all the variables used in this paper and their corresponding definitions. Here $L$ denotes the LDR frame, $H$ denotes the HDR frame, and $T$ denotes the tonemapped HDR. An important point to note here is that $i$ represents the $i^{th}$ frame of the video. So $(i-1)^{th}$ and $(i+1)^{th}$ represents previous frame and the next frame with respect to the $i^{th}$ frame. 

\subsection{Self-Supervised Denoising Network}
\label{denoising_section}

\begin{figure}[!b]
\centering
\includegraphics[width=1\columnwidth]{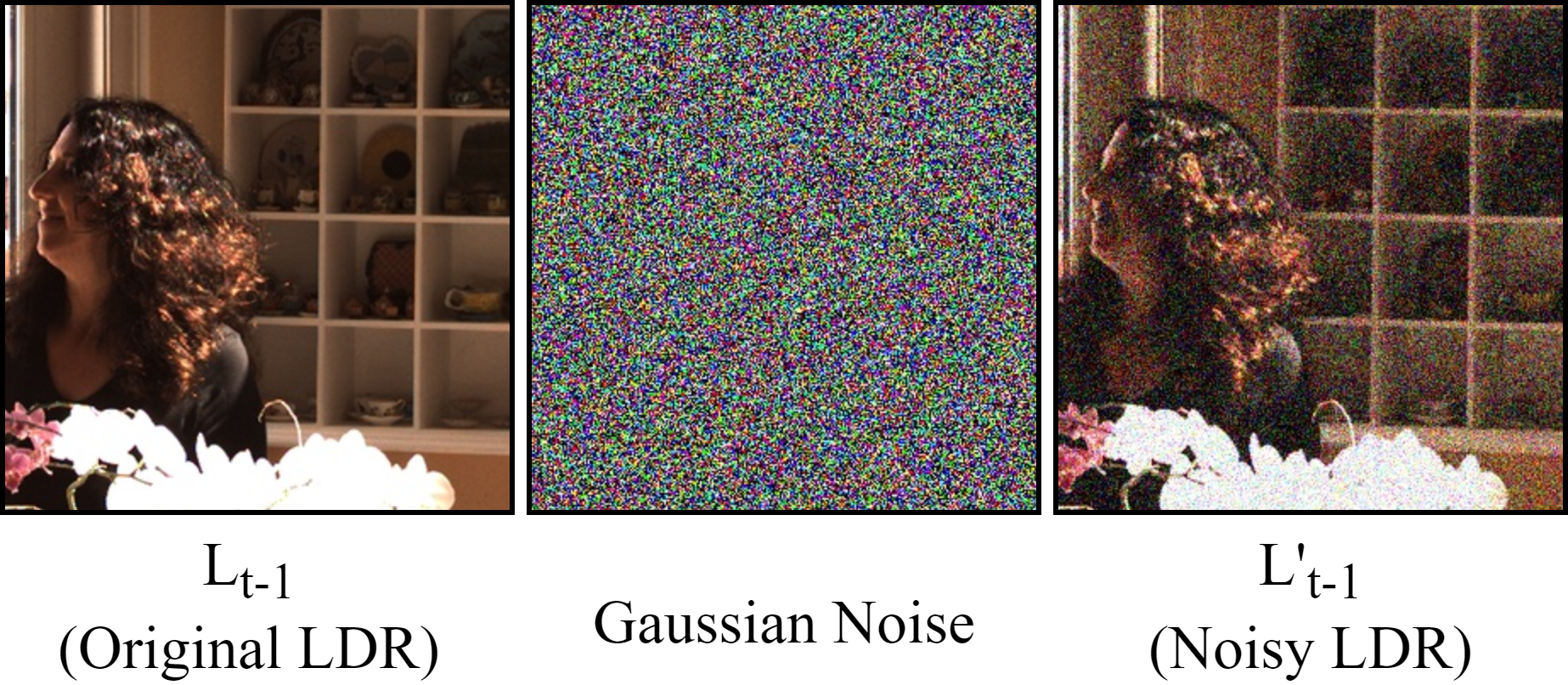}
\caption{Visual comparison of noisy LDR frame $L^{\prime}_{t-1}$ generated by adding gaussian noise using Equation~\ref{noisy_ldr} to the synthetically generated LDR frames $L_{t-1}$}
\label{fig_noise_ldr}
\end{figure}

\begin{figure*}[!t]
\centering
\includegraphics[width=2\columnwidth]{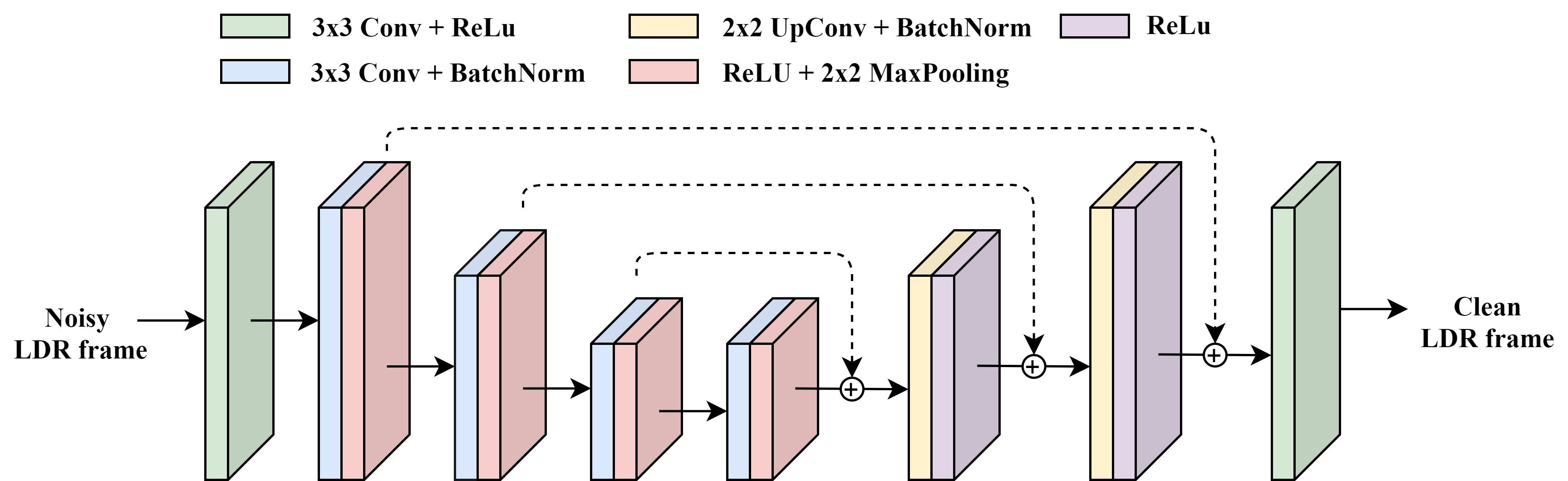}
\caption{Architecture of our Denoising network}
\label{denoise_ldr}
\end{figure*}

Real-world off-the-shelf cameras are prone to capturing noise while recording LDR frames. This noise produces unwanted artifacts in both scenarios, i.e., first while aligning the neighboring frames by computing the optical flow between the frames and secondly on final reconstruction from these aligned frames. In order to reconstruct high-quality HDR frames, we require the corresponding LDR frames to be less noisy. In our method, we incorporate a denoising network that removes such imperfections from the noisy LDR frames. We call our self-supervised denoising blocks as ELDR blocks.  

Self-Supervised paradigm has shown promising results in learning feature representation~\cite{fl,karen}, temporal coherency~\cite{temp}, image denoising~\cite{xu2019noisyasclean}, and many other tasks~\cite{color,ssflow,6869}. Inspired by this, we design a self-supervision based LDR frames denoising network that learns to create clean LDR frames from the noisy LDR video frames.  For all our experimentation, we use Gaussian noise as the perturbation function to generate noisy LDR frames from the synthetic LDR frames as described in Equation~\ref{noisy_ldr}, consistent with previous baselines~\cite{kalantari2019deep}. Figure~\ref{fig_noise_ldr} shows the example of an over-exposed LDR and the corresponding frame after the Gaussian noise addition.  
\begin{equation} \label{noisy_ldr}
\begin{split}
{L^{\prime}}_{i}=L_{i}+N(\mu, \sigma)
\end{split}
\end{equation}

Our LDR frame denoise network consists of a series of convolution and deconvolution operations along with a skip connection between them following a U-Net~\cite{ronneberger2015unet} like structure. Each 2-D convolution operation is followed by a BatchNorm operation, ReLU activation, and a $2 \times 2$ Max-pooling layer to reach the bottleneck representation. Then the bottleneck feature map is upsampled using deconvolution layers. Figure~\ref{denoise_ldr} shows the architecture of our denoising network. In each iteration of the training procedure, we take the LDR frames ($L_i$), and we add perturbation to it according to Equation~\ref{noisy_ldr} and use these perturbed images as the input to the network. The network is trained to extract out the original clean LDR frames from all the added noise. Our proposed method consists of two such denoising networks, each for a different exposure. All the parameters are identical across both networks. The difference is that one network is trained on LDR frames with low exposures and another for LDR frames with high exposures. Formally, we train our denoising blocks using the $L1$ loss given below.
\begin{equation} \label{self_supervised_l1}
\begin{split}
L_{denoise}'=||\widetilde{L}_i - L_i||_1
\end{split}
\end{equation}

\subsection{Flow Network}
% write about integration of flow
In video-to-video synthesis tasks, object movements across frames are known to create temporal artifacts during the reconstruction. Visually incoherent frames with poor temporal coherency is observed if existing image synthesis is directly applied to videos without incorporating the temporal dynamics in the model. We address this by aligning all the input LDR frames with alternating exposure to a reference frame before using it for reconstructing the HDR frames. In order to achieve such an alignment, we first estimate the optical flow~\cite{original_flow} between the consecutive LDR frames having alternating exposures, which is then used to warp the previous frame to the current frame.

Convolution neural network-based optical flow estimation was originally proposed by Dosovitskiy~\emph{et al.}~\cite{first_CNNflownet}, which directly generates a flow field from a pair of images. After this, many works have been proposed on neural network-based optical flow estimation~\cite{pyramid_flow, flownet2, maurer2018proflow}. However, most of these techniques are computationally expensive, and direct application of these methods in our case, would not be scalable for real-time estimation of HDR frames.

Therefore, we use a fine-tuned version of LiteFlowNet~\cite{hui18liteflownet} in our proposed pipeline for optical flow estimation, which outperforms most other neural network-based flow estimation methods both in terms of speed and accuracy. After obtaining the clean version of LDR frames with alternating exposures, we compute the optical flow between these neighboring frames. Originally, LiteFlowNet~\cite{hui18liteflownet} was trained to generate flow-maps between video-frames having similar exposures. Directly using a pre-trained version of this LiteFlowNet~\cite{hui18liteflownet} would result in inconsistencies due to the difference in exposures. In order to utilize this across different exposures, we fine-tune it by leaving it trainable during our end-to-end training procedure after initializing LiteFlowNet~\cite{hui18liteflownet} with the pre-trained version.

\subsection{GAN Based HDR Frame Generation}

\begin{figure*}[!ht]
\centering

\includegraphics[width=2\columnwidth]{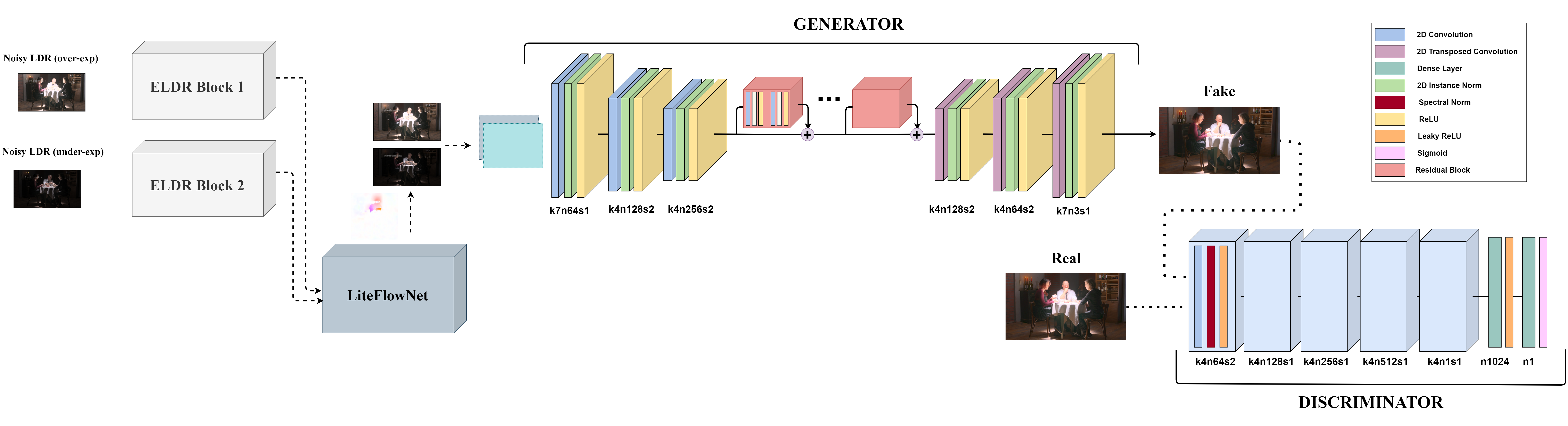}
\caption{Our proposed method for HDR video generation consisting of two Denoising networks, a LiteFlownet~\cite{hui18liteflownet} model, and a final GAN based reconstruction model. Layers in both generator and discriminator can be identified by its color as described in the table on the right side. Each label of the layer follows the convention of k-\textit{kernel size}-n-\textit{number of kernels}-s-\textit{stride size}.}
\label{main_dia}
\end{figure*}

\textbf{Network Architecture.} We adopt the GAN based architecture for our HDR frame reconstruction proposed by~\cite{thasarathan2019automatic}. The proposed generator network is based on encoder-decoder architecture, where the encoder first downsamples the image twice ($H \times W \to H/ 4 \times W/4$). The feature map is then passed through 8 Res-Blocks followed by two upsampling layers. Similar to~\cite{thasarathan2019automatic}, we use the instance norm layer. We warp a convolution layer, an instance-norm layer, and a ReLU activation layer into one basic unit. The Res-Block consists of 2 basic units stacked over each other. The discriminator consists of 5 convolution layers followed by two dense layers. To stabilize the training process, we use the spectral norm. By restricting our discriminative function to 1-Lipschitz, we prevent the gradient uninformativeness problem~\cite{lipschitz}.  Figure~\ref{main_dia} shows our proposed architecture. Let \textit{G} and \textit{D} denotes the generator and discriminator networks. Our generator network takes clean overexposed LDR ($\widetilde{L}_i$) and an underexposed LDR ($\widetilde{L}_{i-1}$) received after flow correction from the denoising networks and generates the current HDR frame ($H_{i}$). 

\begin{equation} \label{generator}
\begin{split}
H_{i}=\textit{G}(\widetilde{L}_{i},\widetilde{L}_{i-1})
\end{split}
\end{equation}

\subsection{Objective Function}

\textbf{Tone Mapping.} Kalantari~\emph{et al.}~\cite{kalantari2019deep} argued that defining loss function in linear HDR domain underestimates the error in darker regions. The solution that they suggested is to convert HDR from linear domain to log domain~\cite{kalantari2017deep,zhang-iccv-17}. Consistent with the previous baselines~\cite{kalantari2019deep}, we also use differentiable  $\mu$-law function transformation denoted by $T$. We compute our loss on tonemapped HDR frames.
\begin{equation} \label{tonemap}
\begin{split}
T_{i}=\frac{log(1+\mu H_i)}{log(1+\mu)}
\end{split}
\end{equation}

\textbf{\boldmath {$L_1$} Loss.} Kalantari~\emph{et al.}~\cite{kalantari2019deep} used $l_1$ loss computed between generated HDR frame and ground truth HDR frame to train the model. The authors also argued that the use of $L_1$ loss promotes sharpness in images as compared to $L_2$ loss. Again, consistent with the previous baselines, we also use $L_1$ loss in our model. 
\begin{equation} \label{l1}
\begin{split}
L_{l_1}= ||T_i - \widetilde{T}_{i} ||_1
\end{split}
\end{equation}

\textbf{Adversarial Objective.} Rather than discriminating with $i^{th}$ ground truth HDR frame, Thasarathan and Nareri~\cite{thasarathan2019automatic} proposed use of previous $(i-1)^{th}$ frame. They argued that using a previous frame for discrimination generates temporally more coherent frames than frames generated using conventional adversarial loss. This adversarial loss reduces flickering artifacts in frames.

\begin{equation} \label{adv}
\begin{split}
L_{adv} & =\mathbb{E}_{(\widetilde{L}_{i},\widetilde{L}_{i-1})}log[\textit{D}(\widetilde{L}_{i},\widetilde{L}_{i-1})] \\ 
& + \mathbb{E}_{\widetilde{L}_{i-1}}log[1-\textit{D}(T_i,\widetilde{T}_{i-1})]
\end{split}
\end{equation}

\textbf{Content and Style Losses.} Frame reconstruction is also accompanied by other visual artifacts like blurriness and color mismatch. We incorporate content and style loss~\cite{sanakoyeu2018styleaware} to minimize visual artifacts. Let $\phi_i$ represent the activated feature map of $j^{th}$ layer of pre-trained VGG-19. For our experiments, we use feature maps of $1^{st}$ to $5^{th}$ layers. We also use style loss to maintain spatial consistency in the generated HDR frame. $\Delta^{\phi}_j$ represents Gram matrix of $j^{th}$ feature map $\phi$. Equation \ref{content} represents content loss and  Equation \ref{style} represents style loss. 
\begin{equation} \label{content}
\begin{split}
L_{content}=\mathbb{E}_{j} \left[ \frac{1}{N_j}||\phi_j(T_i) - \phi_j(\widetilde{T}_i) ||_1 \right]
\end{split}
\end{equation}
\begin{equation} \label{style}
\begin{split}
L_{style}=\mathbb{E}_{j} \left[ ||\Delta^{\phi}_j(T_i) - \Delta^{\phi}_j(\widetilde{T}_i) ||_1 \right]
\end{split}
\end{equation}

For our experiments we use $\lambda_{adv}=5$, $\lambda_{content}=1$, $\lambda_{style}=1000$ and $\lambda_{l_{1}}=30$, Equation \ref{final} represents the overall loss function.

\begin{equation} 
\begin{split}
L_{rec}= \lambda_{adv}L_{adv} + \lambda_{content}L_{content} + \lambda_{style}L_{content} \\+ \lambda_{style}L_{style} +\lambda_{l_{1}}L_{l_{1}}
\end{split}
\label{final}
\end{equation}

\textbf{Temporal Regularization.} Finally, we incorporate explicit regularization for additional temporal stability~\cite{bong,eilertsen_mantiuk_unger_2019} between two consecutive frames, which further helps in reducing blurriness in high motion frames. \textit{W} represents warping function from our flow network and $\alpha$ is our regularization parameter, we set $\alpha=0.3$. 
\begin{equation} 
\begin{split}
L_{reg}=||T_{i} - W({T}_{i-1}) ||_2 
\end{split}
\end{equation}
\begin{equation} 
\begin{split}
L_{total} = \alpha L_{rec} + (1-\alpha)L_{reg} 
\end{split}
\end{equation}

\section{Training Details}
\label{training_det}
In this section, we discuss our training methodology and present the values of hyper-parameters. For all of our experiments, we used $\mu=5000$ for tonemapping from linear to a logarithmic scale. We train our self-supervised denoising networks with $L_{l2}$ loss for 100 epochs. We train our GAN model with $L_{rec}$ loss using a batch-size of 20 for 70 epochs, and then we fine-tune our network with $L_{total}$ using a batch-size of 35 for 15 epochs. The training was performed entirely on a machine with Intel Core i7, 64GB of memory, and a GeForce RTX 2080-Ti GPU. It roughly takes six days to complete the training procedure (both denoising and GAN combined). Note that we freeze the weights for our denoising network before training our GAN model. We optimize our loss objective function using Adam optimizer with a learning rate of $10^{-4}$ for training self-supervised network and $10^{-4}$ for GANs with a batch size of 12. We use Leaky ReLU activation for the discriminator network~\cite{thasarathan2019automatic}, and for the rest of the network, we used ReLU activation. We used the spectral norm in the discriminator to stabilize our training procedure.

\section{Results}
\label{results}
We compare our approach against the method of Kalantari~\emph{et al.}~\cite{kalantari2013patch}, which uses a patch-based mechanism for high dynamic range video generation and against the current state-of-the-art, Kalantari~\emph{et al.}~\cite{kalantari2019deep} which is based on convolutional neural networks (CNNs). We used the publicly available source code for the patch-based method by Kalantari~\emph{et al.}~\cite{kalantari2013patch} and Li~\emph{et al.}~\cite{li2016maximum}. For CNN based approach by Kalantari~\emph{et al.}~\cite{kalantari2019deep}, the authors provided their results on only three scenes from the test set. Both the patch-based mechanism by Kalantari~\emph{et al.}~\cite{kalantari2013patch} and Li~\emph{et al.}~\cite{li2016maximum} takes roughly 1-2 hours for generating each of the frames with a resolution $1280\times 720$. Recent work on CNN based method by Kalantari~\emph{et al.}~\cite{kalantari2019deep} showed that both patch-based method~\cite{kalantari2013patch} and the method of Li~\emph{et al.}~\cite{li2016maximum} produce poor results on different scenes. Thus, visual comparison against patch-based method~\cite{kalantari2013patch} and the method of Li~\emph{et al.}~\cite{li2016maximum} is difficult and superfluous. Moreover, our proposed method has a training mechanism that is consistent with that of the recent CNN based model by Kalantari~\emph{et al.}~\cite{kalantari2019deep} as explained in Section~\ref{training_det}. Therefore, we only compare the visual results against CNN based model of Kalantari~\emph{et al.}~\cite{kalantari2019deep}. 

\subsection{Evaluation Metrics}

\begin{table}[!b]
            \centering
        \begin{tabular}{ccccc} 
        \toprule
         & Kalantari~\cite{kalantari2013patch} & Kalantari~\cite{kalantari2019deep} & Ours \\ 
        [0.5ex] 
        \hline 
        \noalign{\vskip 0.1cm}
        PSNR &38.77&40.67&\textbf{43.35}\\
        SSIM &-&0.78&\textbf{0.83}\\        
        HDR-VDP-2 &62.12&74.15&\textbf{77.19}\\
        \bottomrule
        \end{tabular}
        \caption{Quantitative comparison of our method against the patch based method of Kalantari~\emph{et al.}~\cite{kalantari2013patch} and CNN based method by Kalantari~\emph{et al.}~\cite{kalantari2019deep}.}
        \label{tab:eval}
\end{table}

To evaluate the performance of our end-to-end generative model we use PSNR~\cite{psnr}, SSIM~\cite{ssim} and HDR-VDP-2~\cite{vdp}. Given the ground truth image$(gt)$ and the predicted image$(pred)$, PSNR$(gt, pred)$ is defined as in equation~\ref{psnr} -
\begin{equation}
\label{psnr}
PSNR (gt,pred)=10\log_{10}(255^{2}/MSE(gt,pred))
\end{equation}

where $MSE(gt,pred)$ is mean squared error between the ground truth image and predicted image having a size of $M \times N$ as in equation~\ref{mse} -
\begin{equation}
\label{mse}
MSE(gt,pred)={1\over M \cdot N}\sum_{i=1}^{M}\sum_{j=1}^{N}(gt_{ij}-pred_{ij})^{2}    
\end{equation}

Higher the PSNR value better is the quality of the reconstructed image. The SSIM metric is also a well-known metric for measuring the visual quality of the reconstructed image. SSIM metric takes into account luminance, contrast and structural similarity into account and hence it is highly correlated with the human perception. 
% and is defined as in equation~\ref{ssim} -
% \begin{equation}
% \label{ssim}
% SSIM (gt,pred)=l(gt,pred) \cdot c(gt,pred) \cdot s(gt,pred)    
% \end{equation}

% where,
% \[
% \begin{cases}
% l(gt,pred)={2\mu_{gt}\mu_{pred}+C_{1} \over \mu_{gt}^{2}+\mu_{pred}^{2}+C_{1}}\cr \\ c(gt,pred)={2\sigma_{gt}\sigma_{pred}+C_{2} \over \sigma_{gt}^{2}+\sigma_{pred}^{2}+C_{2}}\cr \\ 
% s(gt,pred)={\sigma_{gt\cdot pred}+C_{3} \over \sigma_{gt}\sigma_{pred}+C_{3}}\cr \\
% \end{cases} 
% \]
% Here the first term $l(gt, pred)$ measures the luminance between the $gt$ image and $pred$ image. The second term $c(gt, pred)$ measures the contrast between the $gt$ and $pred$. While the last term $s(gt, pred)$ measures the correlation coefficient that gives the structure comparison between $gt$ and $pred$.

HDR-VDP-2~\cite{vdp} is also a visual metric that compares the visibility score, i.e., the difference between $gt$ and $pred$ with respect to an average observer and the degradation quality of $pred$ with respect to $gt$ expressed as a mean opinion score.

\subsection{Quantitative Comparison} 
We quantitatively compare our results against the methods of Kang~\emph{et al.}~\cite{khan2006ghost}, the patch-based method of Kalantari~\emph{et al.}~\cite{kalantari2013patch}, and the current state-of-the-art by Kalantari~\emph{et al.}~\cite{kalantari2019deep}. We select frames from the scenes of \uppercase{FISHING LONGSHOT, CAROUSEL FIREWORKS,} and \uppercase{POKER FULLSHOT,} which all comprise of the test set. We extract LDR frames with alternating exposures, as described in previous sections. Each frame has a resolution of $1920\times1080$, but has a wide black border of 10 pixels around them, which we crop on-wards for quantitative comparison.

We evaluate the results on PSNR (Peak signal-to-noise ratio) and SSIM (Structural Similarity Index) in its tone-mapped domain as described in Equation~\ref{tonemap}. To further evaluate the quality of the generated HDR frames, we use HDR-VDP2~\cite{hdr_vdp2}, which is designed specifically to evaluate HDR images and videos. Table~\ref{tab:eval} shows all the values using these metrics computed and averaged across all the frames on test data. It can be seen from Table~\ref{tab:eval} that the proposed method outperforms the other existing approaches with respect to all the considered metrics.

\begin{figure}[!b]
\centering
\includegraphics[width=1\columnwidth]{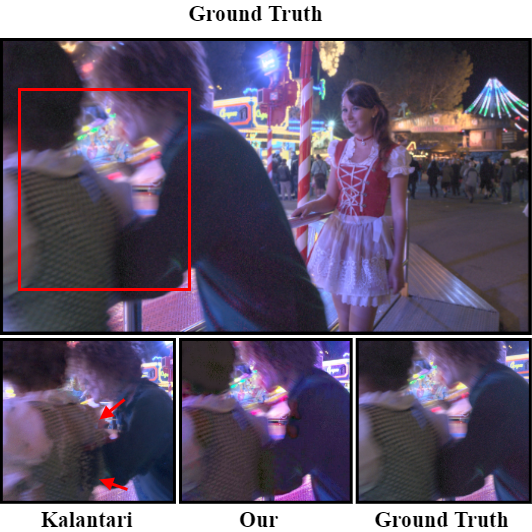}
\caption{Visual Comparison of our generated HDR frames having high motion from \uppercase{CAROUSEL FIREWORKS} scene in test data against Kalantari~\emph{et al.}~\cite{kalantari2019deep}.}
\label{res2_motion1}
\end{figure}

\begin{figure*}[!t]
\centering
\includegraphics[width=2\columnwidth]{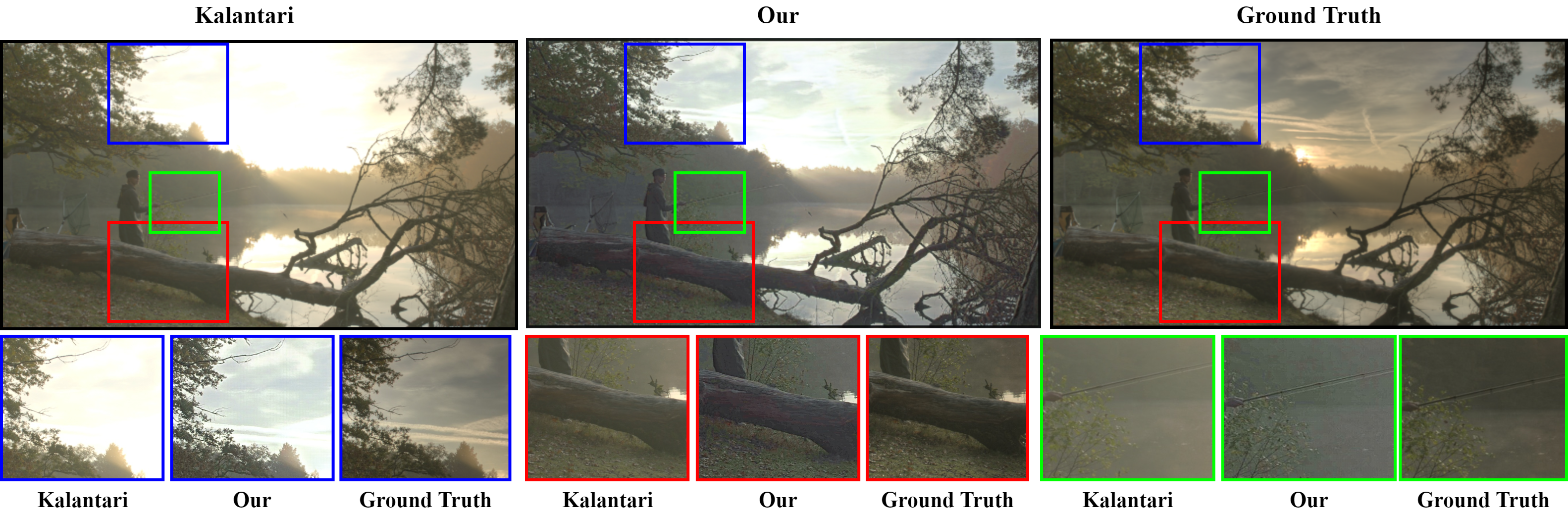}
\caption{Visual Comparison of the our generated HDR frames from a scene of \uppercase{FISHING LONGSHOT} in the test data against Kalantari~\emph{et al.}~\cite{kalantari2019deep}. Identical regions of comparison are all grouped in the same color.}
\label{result_big}
\end{figure*}

\subsection{Visual Comparisons}

\begin{figure}[!b]
\centering
\includegraphics[width=1\columnwidth]{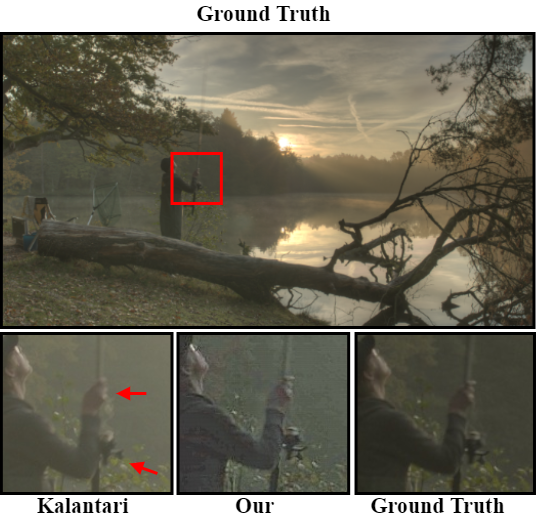}
\caption{Visual Comparison of our generated HDR frames having high motion from \uppercase{FISHING LONGSHOT} scene in test data against Kalantari~\emph{et al.}~\cite{kalantari2019deep}.}
\label{res2_motion2}
\end{figure}

We compare the reconstructed HDR frames of our method, and the CNN based method of Kalantari~\emph{et al.}~\cite{kalantari2019deep} on scenes in the test set. Figure~\ref{result_big} shows a detailed comparison of an HDR frame from a test video scene. The scene shows the \uppercase{FISHING LONGSHOT}, which includes a bright region exposed by the sun (marked in the blue box) and a dark region having very low exposure (marked in red box). It can be clearly observed from the regions bounded by the blue and green color that the proposed method produces frames with much more dynamic range than that of Kalantari~\emph{et al.}~\cite{kalantari2019deep}. On a close observation near the region of the sky (bounded in blue) in Figure~\ref{result_big}, it can be seen that our method reconstructs the details of the clouds, while the one produced by the method of Kalantari~\emph{et al.}~\cite{kalantari2019deep} loses the content and reconstructs an over-exposed frame. Overall, from Figure~\ref{result_big} it can be seen that our method generates HDR frames with much more details in regions of both the high and low exposure areas. 

Figure~\ref{res2_motion1} and Figure~\ref{res2_motion2}  compare our approach against the CNN based method by Kalantari~\emph{et al.}~\cite{kalantari2019deep} on the scenes of \uppercase{CAROUSEL FIREWORKS} and \uppercase{FISHING LONGSHOT}  respectively, which were not part of the training set. We selected those frames to form the video scene with significant motion in-between the adjacent frames i.e. high motion frames. As evident from the frames of \uppercase{CAROUSEL FIREWORKS} in Figure~\ref{res2_motion1}, CNN based method by Kalantari~\emph{et al.}~\cite{kalantari2019deep} generates tearing artifacts in moving parts of the person in the frame. It also produces blurred frames with ghosting artifacts in the scene of \uppercase{FISHING LONGSHOT} in Figure~\ref{res2_motion2} marked by red arrows.

\subsection{Denoising Network}

\begin{figure}[!b]
\centering
\includegraphics[width=1\columnwidth]{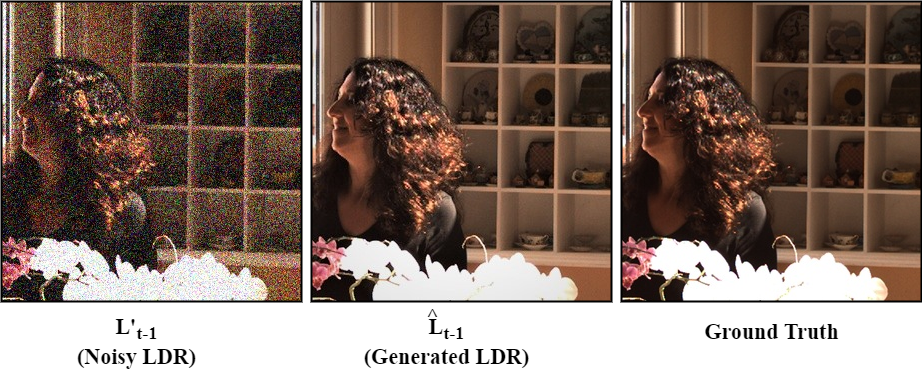}
\caption{Visual comparison of a reconstructed clean LDR frame using denoising network from a noisy LDR frame of a scene used previously in Figure~\ref{fig_noise_ldr}}
\label{fig_noise_res}
\end{figure}

We begin by showing the results of the denoising network (ELDR Blocks), which is trained in a self-supervised manner on the same training set scenes. We extract LDR frames with alternating exposures from the HDR videos, as described in Section~\ref{dataset_section}. For each LDR frame having alternating exposures, we add Gaussian noise with varied signal-to-noise ratios (SNR), which we described in Section~\ref{denoising_section}. Figure~\ref{fig_noise_res} shows the results of a clean LDR frame generated by our denoising network from a noisy LDR frame of a scene, which was earlier described in Figure~\ref{fig_noise_ldr}. In general, we observe that the generated LDR frames have a coherent texture with sharp features as compared to the noisy LDR frames. The use of $L_1$ loss function in the ELDR blocks can be accounted for the above observation.

\section{Ablation Study}

\label{sec:ablation_study}

\begin{figure}[!h]
\centering
\includegraphics[width=1\columnwidth]{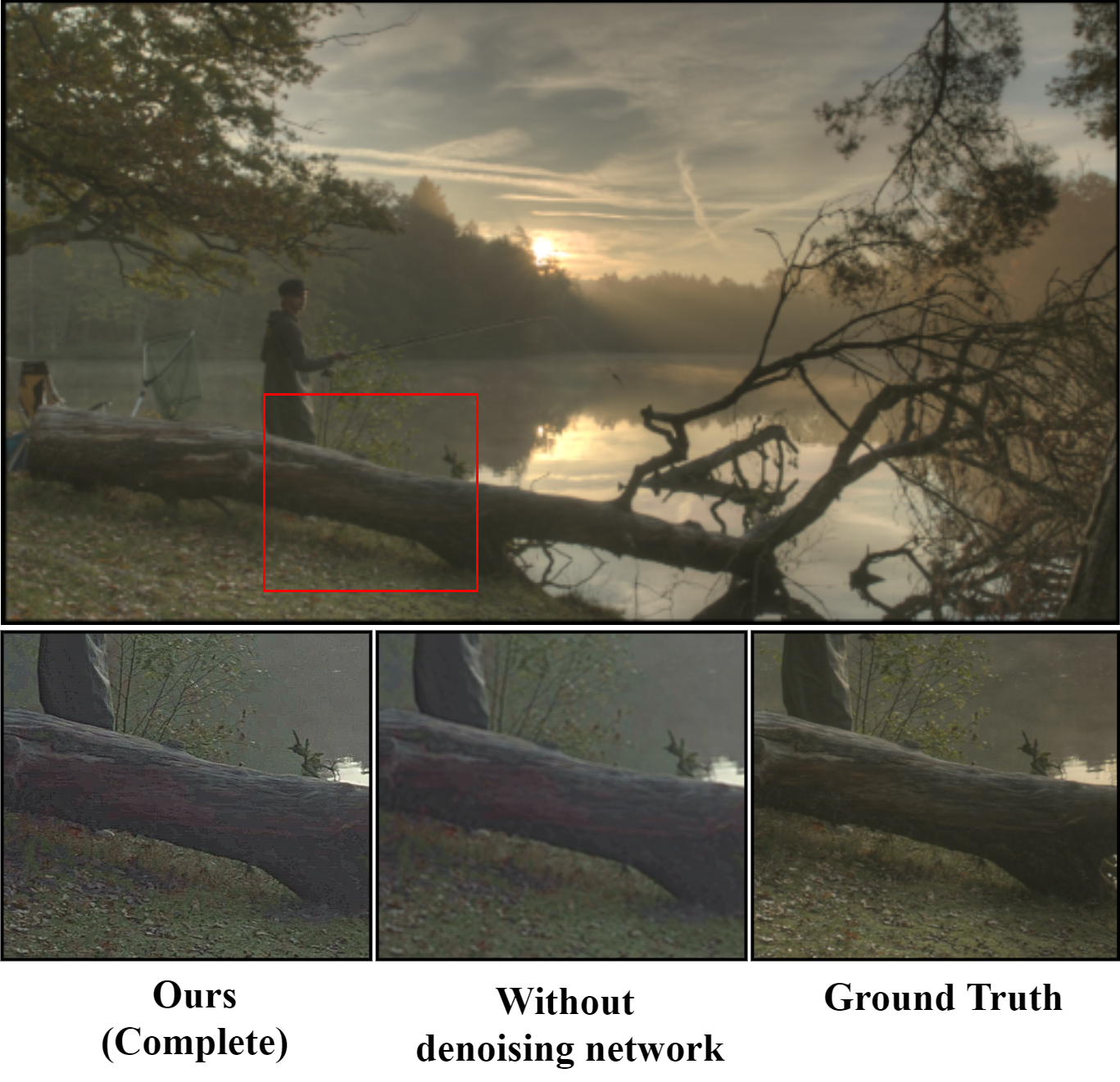}
\caption{Visual comparison of our complete approach to our method without the denoising network. Removing the denoising network leads to reconstruction of HDR frames having poor quality with less details.}
\label{ablation}
\end{figure}

\textbf{Importance of Separate LDR Denoising Blocks.} To study the significance of these ELDR blocks, we remove these blocks from our overall pipeline and re-train the model. It is evident from Table~\ref{tab:ablation} and Figure~\ref{ablation} that both visual quality and metric-wise, removal of denoising network have a significant effect on the performance of our method. We observe a notable drop in the PSNR, SSIM, and HDR-VDP2~\cite{hdr_vdp2} values. We find HDR frames generated directly from noise embedded LDRs to be blurry with less detailed reconstruction as compared to our complete approach, which generates crisp details as shown in Figure~\ref{ablation}. Thus, creating a two-stage network where noise removal and reconstruction of HDR videos are performed separately has shown to perform better than a single network performing both the tasks. Moreover, the generated intermediate clean LDR frames add more interpretability in terms of noise removal to our model as compared to previous baselines~\cite{kalantari2019deep,kalantari2017deep}.

\begin{table}[!h]
            \centering
        \begin{tabular}{ccccc} 
        \toprule
         & Without denosing net & Ours (Complete) \\ 
        \hline 
        \noalign{\vskip 0.1cm}
        PSNR &41.39&\textbf{43.35}\\
        SSIM &0.76&\textbf{0.83}\\        
        HDR-VDP-2 &73.87&\textbf{77.19}\\
        \bottomrule
        \end{tabular}
        % \noalign{\vskip 0.1cm}        
        \caption{Quantitative comparison of our complete approach to our method without denoising network. Removing the denoising network clearly drops the performace on all the mentioned metrics.}
        \label{tab:ablation}
\end{table}%

\section{Conclusion}
\label{conclusion}
In this paper, we proposed a temporally stable GAN-based HDR video reconstruction network that reconstructs HDR videos from LDR sequences with alternating exposures. Our method incorporates a separate LDR denoising network for extracting clean LDR frames, and we showed that creating separate denoising and reconstruction network outperforms a single network that performs both the tasks. We first align the neighboring alternating exposure frames using the LiteFlownet~\cite{hui18liteflownet} to generate temporally coherent frames. Training our model over a joint objective consisting of $L_1$ loss, style-aware content losses~\cite{sanakoyeu2018styleaware} and augmented GAN loss~\cite{thasarathan2019automatic} helped in minimizing the visual artifacts. Further, we fine-tune our model on a temporal stability based regularization term to further reduce the tearing and ghosting artifacts due to temporal incoherence. 
We perform all our experimentation consistent with the previous baselines, and we demonstrate that our method outperforms the previous baselines both visually and metric-wise. 

We believe that there is a great scope for further improvement in terms of better colors, more dynamic range, and in overall visual quality. In the future, we would like to further test our proposed method on larger datasets of HDR videos as and when they become available. 

%%
%% The acknowledgments section is defined using the "acks" environment
%% (and NOT an unnumbered section). This ensures the proper
%% identification of the section in the article metadata, and the
%% consistent spelling of the heading.
\begin{acks}
We would like to thank Science and Engineering Research Board (SERB) Core Research Grant for supporting our work.
\end{acks}

\bibliographystyle{ACM-Reference-Format}
\bibliography{ICVGIP21-CameraReady-Template}

%%% -*-BibTeX-*-
%%% Do NOT edit. File created by BibTeX with style
%%% ACM-Reference-Format-Journals [18-Jan-2012].

\begin{thebibliography}{59}

%%% ====================================================================
%%% NOTE TO THE USER: you can override these defaults by providing
%%% customized versions of any of these macros before the \bibliography
%%% command.  Each of them MUST provide its own final punctuation,
%%% except for \shownote{}, \showDOI{}, and \showURL{}.  The latter two
%%% do not use final punctuation, in order to avoid confusing it with
%%% the Web address.
%%%
%%% To suppress output of a particular field, define its macro to expand
%%% to an empty string, or better, \unskip, like this:
%%%
%%% \newcommand{\showDOI}[1]{\unskip}   % LaTeX syntax
%%%
%%% \def \showDOI #1{\unskip}           % plain TeX syntax
%%%
%%% ====================================================================

\ifx \showCODEN    \undefined \def \showCODEN     #1{\unskip}     \fi
\ifx \showDOI      \undefined \def \showDOI       #1{#1}\fi
\ifx \showISBNx    \undefined \def \showISBNx     #1{\unskip}     \fi
\ifx \showISBNxiii \undefined \def \showISBNxiii  #1{\unskip}     \fi
\ifx \showISSN     \undefined \def \showISSN      #1{\unskip}     \fi
\ifx \showLCCN     \undefined \def \showLCCN      #1{\unskip}     \fi
\ifx \shownote     \undefined \def \shownote      #1{#1}          \fi
\ifx \showarticletitle \undefined \def \showarticletitle #1{#1}   \fi
\ifx \showURL      \undefined \def \showURL       {\relax}        \fi
% The following commands are used for tagged output and should be
% invisible to TeX
\providecommand\bibfield[2]{#2}
\providecommand\bibinfo[2]{#2}
\providecommand\natexlab[1]{#1}
\providecommand\showeprint[2][]{arXiv:#2}

\bibitem[\protect\citeauthoryear{Bonneel, Tompkin, Sunkavalli, Sun, Paris, and
  Pfister}{Bonneel et~al\mbox{.}}{2015}]%
        {bong}
\bibfield{author}{\bibinfo{person}{Nicolas Bonneel}, \bibinfo{person}{James
  Tompkin}, \bibinfo{person}{Kalyan Sunkavalli}, \bibinfo{person}{Deqing Sun},
  \bibinfo{person}{Sylvain Paris}, {and} \bibinfo{person}{Hanspeter Pfister}.}
  \bibinfo{year}{2015}\natexlab{}.
\newblock \showarticletitle{Blind Video Temporal Consistency}.
\newblock \bibinfo{journal}{\emph{ACM Trans. Graph.}} \bibinfo{volume}{34},
  \bibinfo{number}{6}, Article \bibinfo{articleno}{196} (\bibinfo{date}{Oct.}
  \bibinfo{year}{2015}), \bibinfo{numpages}{9}~pages.
\newblock
\showISSN{0730-0301}
\urldef\tempurl%
\url{https://doi.org/10.1145/2816795.2818107}
\showDOI{\tempurl}


\bibitem[\protect\citeauthoryear{Brajovic and Kanade}{Brajovic and
  Kanade}{1996}]%
        {brajovic1996sorting}
\bibfield{author}{\bibinfo{person}{Vladimir Brajovic} {and}
  \bibinfo{person}{Takeo Kanade}.} \bibinfo{year}{1996}\natexlab{}.
\newblock \showarticletitle{A sorting image sensor: An example of massively
  parallel intensity-to-time processing for low-latency computational sensors}.
  In \bibinfo{booktitle}{\emph{Proceedings of IEEE International Conference on
  Robotics and Automation}}, Vol.~\bibinfo{volume}{2}.
  \bibinfo{publisher}{IEEE}, \bibinfo{pages}{1638--1643}.
\newblock


\bibitem[\protect\citeauthoryear{Debevec and Malik}{Debevec and Malik}{1997}]%
        {debevec1997recovering}
\bibfield{author}{\bibinfo{person}{PE Debevec} {and} \bibinfo{person}{J
  Malik}.} \bibinfo{year}{1997}\natexlab{}.
\newblock \showarticletitle{Recovering high dynamic range radiance maps from
  photographs: Proceedings of the 24th Annual Conference on Computer Graphics
  and Interactive Techniques}.
\newblock \bibinfo{journal}{\emph{Los Angeles, USA: SIGGRAPH}}
  (\bibinfo{year}{1997}).
\newblock


\bibitem[\protect\citeauthoryear{Debevec and Malik}{Debevec and Malik}{2008}]%
        {debevec2008recovering}
\bibfield{author}{\bibinfo{person}{Paul~E Debevec} {and}
  \bibinfo{person}{Jitendra Malik}.} \bibinfo{year}{2008}\natexlab{}.
\newblock \showarticletitle{Recovering high dynamic range radiance maps from
  photographs}.
\newblock In \bibinfo{booktitle}{\emph{ACM SIGGRAPH 2008 classes}}.
  \bibinfo{pages}{1--10}.
\newblock


\bibitem[\protect\citeauthoryear{Eilertsen, Kronander, Denes, Mantiuk, and
  Unger}{Eilertsen et~al\mbox{.}}{2017}]%
        {eilertsen2017hdr}
\bibfield{author}{\bibinfo{person}{Gabriel Eilertsen}, \bibinfo{person}{Joel
  Kronander}, \bibinfo{person}{Gyorgy Denes}, \bibinfo{person}{Rafa{\l}~K
  Mantiuk}, {and} \bibinfo{person}{Jonas Unger}.}
  \bibinfo{year}{2017}\natexlab{}.
\newblock \showarticletitle{HDR image reconstruction from a single exposure
  using deep CNNs}.
\newblock \bibinfo{journal}{\emph{ACM Transactions on Graphics (TOG)}}
  \bibinfo{volume}{36}, \bibinfo{number}{6} (\bibinfo{year}{2017}),
  \bibinfo{pages}{1--15}.
\newblock


\bibitem[\protect\citeauthoryear{Eilertsen, Mantiuk, and Unger}{Eilertsen
  et~al\mbox{.}}{2019}]%
        {eilertsen_mantiuk_unger_2019}
\bibfield{author}{\bibinfo{person}{Gabriel Eilertsen},
  \bibinfo{person}{Rafal~K. Mantiuk}, {and} \bibinfo{person}{Jonas Unger}.}
  \bibinfo{year}{2019}\natexlab{}.
\newblock \showarticletitle{Single-Frame Regularization for Temporally Stable
  CNNs}.
\newblock \bibinfo{journal}{\emph{2019 IEEE/CVF Conference on Computer Vision
  and Pattern Recognition (CVPR)}} (\bibinfo{year}{2019}).
\newblock
\urldef\tempurl%
\url{https://doi.org/10.1109/cvpr.2019.01143}
\showDOI{\tempurl}


\bibitem[\protect\citeauthoryear{Fischer, Dosovitskiy, Ilg, Häusser,
  Hazırbaş, Golkov, van~der Smagt, Cremers, and Brox}{Fischer
  et~al\mbox{.}}{2015}]%
        {first_CNNflownet}
\bibfield{author}{\bibinfo{person}{Philipp Fischer}, \bibinfo{person}{Alexey
  Dosovitskiy}, \bibinfo{person}{Eddy Ilg}, \bibinfo{person}{Philip Häusser},
  \bibinfo{person}{Caner Hazırbaş}, \bibinfo{person}{Vladimir Golkov},
  \bibinfo{person}{Patrick van~der Smagt}, \bibinfo{person}{Daniel Cremers},
  {and} \bibinfo{person}{Thomas Brox}.} \bibinfo{year}{2015}\natexlab{}.
\newblock \bibinfo{title}{FlowNet: Learning Optical Flow with Convolutional
  Networks}.
\newblock
\newblock
\showeprint[arxiv]{1504.06852}~[cs.CV]


\bibitem[\protect\citeauthoryear{Froehlich, Grandinetti, Eberhardt, Walter,
  Schilling, and Brendel}{Froehlich et~al\mbox{.}}{2014}]%
        {froehlich2014creating}
\bibfield{author}{\bibinfo{person}{Jan Froehlich}, \bibinfo{person}{Stefan
  Grandinetti}, \bibinfo{person}{Bernd Eberhardt}, \bibinfo{person}{Simon
  Walter}, \bibinfo{person}{Andreas Schilling}, {and} \bibinfo{person}{Harald
  Brendel}.} \bibinfo{year}{2014}\natexlab{}.
\newblock \showarticletitle{Creating cinematic wide gamut HDR-video for the
  evaluation of tone mapping operators and HDR-displays}. In
  \bibinfo{booktitle}{\emph{Digital Photography X}},
  Vol.~\bibinfo{volume}{9023}. International Society for Optics and Photonics,
  \bibinfo{pages}{90230X}.
\newblock


\bibitem[\protect\citeauthoryear{Gallo, Troccoli, Hu, Pulli, and Kautz}{Gallo
  et~al\mbox{.}}{2015}]%
        {gallo2015locally}
\bibfield{author}{\bibinfo{person}{Orazio Gallo}, \bibinfo{person}{Alejandro
  Troccoli}, \bibinfo{person}{Jun Hu}, \bibinfo{person}{Kari Pulli}, {and}
  \bibinfo{person}{Jan Kautz}.} \bibinfo{year}{2015}\natexlab{}.
\newblock \showarticletitle{Locally non-rigid registration for mobile HDR
  photography}. In \bibinfo{booktitle}{\emph{Proceedings of the IEEE Conference
  on Computer Vision and Pattern Recognition Workshops}}.
  \bibinfo{pages}{49--56}.
\newblock


\bibitem[\protect\citeauthoryear{Gryaditskaya, Pouli, Reinhard, Myszkowski, and
  Seidel}{Gryaditskaya et~al\mbox{.}}{2015}]%
        {gryaditskaya2015motion}
\bibfield{author}{\bibinfo{person}{Yulia Gryaditskaya}, \bibinfo{person}{Tania
  Pouli}, \bibinfo{person}{Erik Reinhard}, \bibinfo{person}{Karol Myszkowski},
  {and} \bibinfo{person}{Hans-Peter Seidel}.} \bibinfo{year}{2015}\natexlab{}.
\newblock \showarticletitle{Motion aware exposure bracketing for HDR video}. In
  \bibinfo{booktitle}{\emph{Computer Graphics Forum}},
  Vol.~\bibinfo{volume}{34}. Wiley Online Library, \bibinfo{pages}{119--130}.
\newblock


\bibitem[\protect\citeauthoryear{Hasinoff, Sharlet, Geiss, Adams, Barron,
  Kainz, Chen, and Levoy}{Hasinoff et~al\mbox{.}}{2016}]%
        {hasinoff2016burst}
\bibfield{author}{\bibinfo{person}{Samuel~W Hasinoff}, \bibinfo{person}{Dillon
  Sharlet}, \bibinfo{person}{Ryan Geiss}, \bibinfo{person}{Andrew Adams},
  \bibinfo{person}{Jonathan~T Barron}, \bibinfo{person}{Florian Kainz},
  \bibinfo{person}{Jiawen Chen}, {and} \bibinfo{person}{Marc Levoy}.}
  \bibinfo{year}{2016}\natexlab{}.
\newblock \showarticletitle{Burst photography for high dynamic range and
  low-light imaging on mobile cameras}.
\newblock \bibinfo{journal}{\emph{ACM Transactions on Graphics (TOG)}}
  \bibinfo{volume}{35}, \bibinfo{number}{6} (\bibinfo{year}{2016}),
  \bibinfo{pages}{1--12}.
\newblock


\bibitem[\protect\citeauthoryear{{Horé} and {Ziou}}{{Horé} and
  {Ziou}}{2010}]%
        {psnr}
\bibfield{author}{\bibinfo{person}{A. {Horé}} {and} \bibinfo{person}{D.
  {Ziou}}.} \bibinfo{year}{2010}\natexlab{}.
\newblock \showarticletitle{Image Quality Metrics: PSNR vs. SSIM}. In
  \bibinfo{booktitle}{\emph{2010 20th International Conference on Pattern
  Recognition}}. \bibinfo{pages}{2366--2369}.
\newblock
\urldef\tempurl%
\url{https://doi.org/10.1109/ICPR.2010.579}
\showDOI{\tempurl}


\bibitem[\protect\citeauthoryear{Hu, Gallo, Pulli, and Sun}{Hu
  et~al\mbox{.}}{2013}]%
        {hu2013hdr}
\bibfield{author}{\bibinfo{person}{Jun Hu}, \bibinfo{person}{Orazio Gallo},
  \bibinfo{person}{Kari Pulli}, {and} \bibinfo{person}{Xiaobai Sun}.}
  \bibinfo{year}{2013}\natexlab{}.
\newblock \showarticletitle{HDR deghosting: How to deal with saturation?}. In
  \bibinfo{booktitle}{\emph{Proceedings of the IEEE Conference on Computer
  Vision and Pattern Recognition}}. \bibinfo{pages}{1163--1170}.
\newblock


\bibitem[\protect\citeauthoryear{Hui, Tang, and Loy}{Hui et~al\mbox{.}}{2018}]%
        {hui18liteflownet}
\bibfield{author}{\bibinfo{person}{Tak-Wai Hui}, \bibinfo{person}{Xiaoou Tang},
  {and} \bibinfo{person}{Chen~Change Loy}.} \bibinfo{year}{2018}\natexlab{}.
\newblock \showarticletitle{LiteFlowNet: A Lightweight Convolutional Neural
  Network for Optical Flow Estimation}. In
  \bibinfo{booktitle}{\emph{Proceedings of IEEE Conference on Computer Vision
  and Pattern Recognition (CVPR)}}. \bibinfo{pages}{8981--8989}.
\newblock


\bibitem[\protect\citeauthoryear{Ilg, Mayer, Saikia, Keuper, Dosovitskiy, and
  Brox}{Ilg et~al\mbox{.}}{2017}]%
        {flownet2}
\bibfield{author}{\bibinfo{person}{E. Ilg}, \bibinfo{person}{N. Mayer},
  \bibinfo{person}{T. Saikia}, \bibinfo{person}{M. Keuper}, \bibinfo{person}{A.
  Dosovitskiy}, {and} \bibinfo{person}{T. Brox}.}
  \bibinfo{year}{2017}\natexlab{}.
\newblock \showarticletitle{FlowNet 2.0: Evolution of Optical Flow Estimation
  with Deep Networks}. In \bibinfo{booktitle}{\emph{IEEE Conference on Computer
  Vision and Pattern Recognition (CVPR)}}.
\newblock
\urldef\tempurl%
\url{http://lmb.informatik.uni-freiburg.de//Publications/2017/IMKDB17}
\showURL{%
\tempurl}


\bibitem[\protect\citeauthoryear{Jenni and Favaro}{Jenni and Favaro}{2018}]%
        {fl}
\bibfield{author}{\bibinfo{person}{Simon Jenni} {and} \bibinfo{person}{Paolo
  Favaro}.} \bibinfo{year}{2018}\natexlab{}.
\newblock \showarticletitle{Self-Supervised Feature Learning by Learning to
  Spot Artifacts}.
\newblock \bibinfo{journal}{\emph{CoRR}}  \bibinfo{volume}{abs/1806.05024}
  (\bibinfo{year}{2018}).
\newblock
\showeprint[arxiv]{1806.05024}
\urldef\tempurl%
\url{http://arxiv.org/abs/1806.05024}
\showURL{%
\tempurl}


\bibitem[\protect\citeauthoryear{Johnson, Alahi, and Fei-Fei}{Johnson
  et~al\mbox{.}}{2016}]%
        {Johnson2016Perceptual}
\bibfield{author}{\bibinfo{person}{Justin Johnson}, \bibinfo{person}{Alexandre
  Alahi}, {and} \bibinfo{person}{Li Fei-Fei}.} \bibinfo{year}{2016}\natexlab{}.
\newblock \showarticletitle{Perceptual losses for real-time style transfer and
  super-resolution}. In \bibinfo{booktitle}{\emph{European conference on
  computer vision}}. Springer, \bibinfo{pages}{694--711}.
\newblock


\bibitem[\protect\citeauthoryear{Kalantari and Ramamoorthi}{Kalantari and
  Ramamoorthi}{2017}]%
        {kalantari2017deep}
\bibfield{author}{\bibinfo{person}{Nima~Khademi Kalantari} {and}
  \bibinfo{person}{Ravi Ramamoorthi}.} \bibinfo{year}{2017}\natexlab{}.
\newblock \showarticletitle{Deep high dynamic range imaging of dynamic scenes.}
\newblock \bibinfo{journal}{\emph{ACM Trans. Graph.}} \bibinfo{volume}{36},
  \bibinfo{number}{4} (\bibinfo{year}{2017}), \bibinfo{pages}{144--1}.
\newblock


\bibitem[\protect\citeauthoryear{Kalantari and Ramamoorthi}{Kalantari and
  Ramamoorthi}{2019}]%
        {kalantari2019deep}
\bibfield{author}{\bibinfo{person}{Nima~Khademi Kalantari} {and}
  \bibinfo{person}{Ravi Ramamoorthi}.} \bibinfo{year}{2019}\natexlab{}.
\newblock \showarticletitle{Deep HDR Video from Sequences with Alternating
  Exposures}. In \bibinfo{booktitle}{\emph{Computer Graphics Forum}},
  Vol.~\bibinfo{volume}{38}. Wiley Online Library, \bibinfo{pages}{193--205}.
\newblock


\bibitem[\protect\citeauthoryear{Kalantari, Shechtman, Barnes, Darabi, Goldman,
  and Sen}{Kalantari et~al\mbox{.}}{2013}]%
        {kalantari2013patch}
\bibfield{author}{\bibinfo{person}{Nima~Khademi Kalantari},
  \bibinfo{person}{Eli Shechtman}, \bibinfo{person}{Connelly Barnes},
  \bibinfo{person}{Soheil Darabi}, \bibinfo{person}{Dan~B Goldman}, {and}
  \bibinfo{person}{Pradeep Sen}.} \bibinfo{year}{2013}\natexlab{}.
\newblock \showarticletitle{Patch-based high dynamic range video.}
\newblock \bibinfo{journal}{\emph{ACM Trans. Graph.}} \bibinfo{volume}{32},
  \bibinfo{number}{6} (\bibinfo{year}{2013}), \bibinfo{pages}{202--1}.
\newblock


\bibitem[\protect\citeauthoryear{Kang, Uyttendaele, Winder, and Szeliski}{Kang
  et~al\mbox{.}}{2003}]%
        {kang2003high}
\bibfield{author}{\bibinfo{person}{Sing~Bing Kang}, \bibinfo{person}{Matthew
  Uyttendaele}, \bibinfo{person}{Simon Winder}, {and} \bibinfo{person}{Richard
  Szeliski}.} \bibinfo{year}{2003}\natexlab{}.
\newblock \showarticletitle{High dynamic range video}.
\newblock \bibinfo{journal}{\emph{ACM Transactions on Graphics (TOG)}}
  \bibinfo{volume}{22}, \bibinfo{number}{3} (\bibinfo{year}{2003}),
  \bibinfo{pages}{319--325}.
\newblock


\bibitem[\protect\citeauthoryear{Khan, Akyuz, and Reinhard}{Khan
  et~al\mbox{.}}{2006}]%
        {khan2006ghost}
\bibfield{author}{\bibinfo{person}{Erum~Arif Khan}, \bibinfo{person}{Ahmet~Oguz
  Akyuz}, {and} \bibinfo{person}{Erik Reinhard}.}
  \bibinfo{year}{2006}\natexlab{}.
\newblock \showarticletitle{Ghost removal in high dynamic range images}. In
  \bibinfo{booktitle}{\emph{2006 International Conference on Image
  Processing}}. IEEE, \bibinfo{pages}{2005--2008}.
\newblock


\bibitem[\protect\citeauthoryear{Kronander, Gustavson, Bonnet, and
  Unger}{Kronander et~al\mbox{.}}{2013}]%
        {kronander2013unified}
\bibfield{author}{\bibinfo{person}{Joel Kronander}, \bibinfo{person}{Stefan
  Gustavson}, \bibinfo{person}{Gerhard Bonnet}, {and} \bibinfo{person}{Jonas
  Unger}.} \bibinfo{year}{2013}\natexlab{}.
\newblock \showarticletitle{Unified HDR reconstruction from raw CFA data}. In
  \bibinfo{booktitle}{\emph{IEEE international conference on computational
  photography (ICCP)}}. IEEE, \bibinfo{pages}{1--9}.
\newblock


\bibitem[\protect\citeauthoryear{Kronander, Gustavson, Bonnet, Ynnerman, and
  Unger}{Kronander et~al\mbox{.}}{2014}]%
        {kronander2014unified}
\bibfield{author}{\bibinfo{person}{Joel Kronander}, \bibinfo{person}{Stefan
  Gustavson}, \bibinfo{person}{Gerhard Bonnet}, \bibinfo{person}{Anders
  Ynnerman}, {and} \bibinfo{person}{Jonas Unger}.}
  \bibinfo{year}{2014}\natexlab{}.
\newblock \showarticletitle{A unified framework for multi-sensor HDR video
  reconstruction}.
\newblock \bibinfo{journal}{\emph{Signal Processing: Image Communication}}
  \bibinfo{volume}{29}, \bibinfo{number}{2} (\bibinfo{year}{2014}),
  \bibinfo{pages}{203--215}.
\newblock


\bibitem[\protect\citeauthoryear{Lai and Xie}{Lai and Xie}{2019}]%
        {ssflow}
\bibfield{author}{\bibinfo{person}{Zihang Lai} {and} \bibinfo{person}{Weidi
  Xie}.} \bibinfo{year}{2019}\natexlab{}.
\newblock \showarticletitle{Self-supervised Learning for Video Correspondence
  Flow}.
\newblock \bibinfo{journal}{\emph{CoRR}}  \bibinfo{volume}{abs/1905.00875}
  (\bibinfo{year}{2019}).
\newblock
\showeprint[arxiv]{1905.00875}
\urldef\tempurl%
\url{http://arxiv.org/abs/1905.00875}
\showURL{%
\tempurl}


\bibitem[\protect\citeauthoryear{Laine, Karras, Lehtinen, and Aila}{Laine
  et~al\mbox{.}}{2019}]%
        {6869}
\bibfield{author}{\bibinfo{person}{Samuli Laine}, \bibinfo{person}{Tero
  Karras}, \bibinfo{person}{Jaakko Lehtinen}, {and} \bibinfo{person}{Timo
  Aila}.} \bibinfo{year}{2019}\natexlab{}.
\newblock \showarticletitle{High-Quality Self-Supervised Deep Image Denoising}.
\newblock In \bibinfo{booktitle}{\emph{Advances in Neural Information
  Processing Systems 32}}, \bibfield{editor}{\bibinfo{person}{H.~Wallach},
  \bibinfo{person}{H.~Larochelle}, \bibinfo{person}{A.~Beygelzimer},
  \bibinfo{person}{F.~d\textquotesingle Alch\'{e}-Buc},
  \bibinfo{person}{E.~Fox}, {and} \bibinfo{person}{R.~Garnett}} (Eds.).
  \bibinfo{publisher}{Curran Associates, Inc.}, \bibinfo{pages}{6970--6980}.
\newblock
\urldef\tempurl%
\url{http://papers.nips.cc/paper/8920-high-quality-self-supervised-deep-image-denoising.pdf}
\showURL{%
\tempurl}


\bibitem[\protect\citeauthoryear{Li, Lee, and Monga}{Li et~al\mbox{.}}{2016}]%
        {li2016maximum}
\bibfield{author}{\bibinfo{person}{Yuelong Li}, \bibinfo{person}{Chul Lee},
  {and} \bibinfo{person}{Vishal Monga}.} \bibinfo{year}{2016}\natexlab{}.
\newblock \showarticletitle{A maximum a posteriori estimation framework for
  robust high dynamic range video synthesis}.
\newblock \bibinfo{journal}{\emph{IEEE Transactions on Image Processing}}
  \bibinfo{volume}{26}, \bibinfo{number}{3} (\bibinfo{year}{2016}),
  \bibinfo{pages}{1143--1157}.
\newblock


\bibitem[\protect\citeauthoryear{Liu, Yuan, Tang, Uyttendaele, and Sun}{Liu
  et~al\mbox{.}}{2014}]%
        {liu2014fast}
\bibfield{author}{\bibinfo{person}{Ziwei Liu}, \bibinfo{person}{Lu Yuan},
  \bibinfo{person}{Xiaoou Tang}, \bibinfo{person}{Matt Uyttendaele}, {and}
  \bibinfo{person}{Jian Sun}.} \bibinfo{year}{2014}\natexlab{}.
\newblock \showarticletitle{Fast burst images denoising}.
\newblock \bibinfo{journal}{\emph{ACM Transactions on Graphics (TOG)}}
  \bibinfo{volume}{33}, \bibinfo{number}{6} (\bibinfo{year}{2014}),
  \bibinfo{pages}{1--9}.
\newblock


\bibitem[\protect\citeauthoryear{Lucas and Kanade}{Lucas and Kanade}{1981}]%
        {original_flow}
\bibfield{author}{\bibinfo{person}{Bruce~D. Lucas} {and} \bibinfo{person}{Takeo
  Kanade}.} \bibinfo{year}{1981}\natexlab{}.
\newblock \showarticletitle{An Iterative Image Registration Technique with an
  Application to Stereo Vision}. In \bibinfo{booktitle}{\emph{Proceedings of
  the 7th International Joint Conference on Artificial Intelligence - Volume
  2}} (Vancouver, BC, Canada) \emph{(\bibinfo{series}{IJCAI'81})}.
  \bibinfo{publisher}{Morgan Kaufmann Publishers Inc.}, \bibinfo{address}{San
  Francisco, CA, USA}, \bibinfo{pages}{674–679}.
\newblock


\bibitem[\protect\citeauthoryear{Ma, Li, Yong, Wang, Meng, and Zhang}{Ma
  et~al\mbox{.}}{2017}]%
        {ma2017robust}
\bibfield{author}{\bibinfo{person}{Kede Ma}, \bibinfo{person}{Hui Li},
  \bibinfo{person}{Hongwei Yong}, \bibinfo{person}{Zhou Wang},
  \bibinfo{person}{Deyu Meng}, {and} \bibinfo{person}{Lei Zhang}.}
  \bibinfo{year}{2017}\natexlab{}.
\newblock \showarticletitle{Robust multi-exposure image fusion: A structural
  patch decomposition approach}.
\newblock \bibinfo{journal}{\emph{IEEE Transactions on Image Processing}}
  \bibinfo{volume}{26}, \bibinfo{number}{5} (\bibinfo{year}{2017}),
  \bibinfo{pages}{2519--2532}.
\newblock


\bibitem[\protect\citeauthoryear{Mangiat and Gibson}{Mangiat and
  Gibson}{2011}]%
        {mangiat2011spatially}
\bibfield{author}{\bibinfo{person}{Stephen Mangiat} {and}
  \bibinfo{person}{Jerry Gibson}.} \bibinfo{year}{2011}\natexlab{}.
\newblock \showarticletitle{Spatially adaptive filtering for registration
  artifact removal in HDR video}. In \bibinfo{booktitle}{\emph{2011 18th IEEE
  International Conference on Image Processing}}. IEEE,
  \bibinfo{pages}{1317--1320}.
\newblock


\bibitem[\protect\citeauthoryear{Mann, Manders, and Fung}{Mann
  et~al\mbox{.}}{2002}]%
        {mann2002painting}
\bibfield{author}{\bibinfo{person}{Steve Mann}, \bibinfo{person}{Corey
  Manders}, {and} \bibinfo{person}{James Fung}.}
  \bibinfo{year}{2002}\natexlab{}.
\newblock \showarticletitle{Painting with looks: Photographic images from video
  using quantimetric processing}. In \bibinfo{booktitle}{\emph{Proceedings of
  the tenth ACM international conference on Multimedia}}.
  \bibinfo{pages}{117--126}.
\newblock


\bibitem[\protect\citeauthoryear{Mann and Picard}{Mann and Picard}{1994}]%
        {mann1994beingundigital}
\bibfield{author}{\bibinfo{person}{S Mann} {and} \bibinfo{person}{R Picard}.}
  \bibinfo{year}{1994}\natexlab{}.
\newblock \showarticletitle{Being undigital with digital cameras}.
\newblock \bibinfo{journal}{\emph{MIT Media Lab Perceptual}}
  \bibinfo{volume}{1} (\bibinfo{year}{1994}), \bibinfo{pages}{2}.
\newblock


\bibitem[\protect\citeauthoryear{Mantiuk, Kim, Rempel, and Heidrich}{Mantiuk
  et~al\mbox{.}}{2011a}]%
        {vdp}
\bibfield{author}{\bibinfo{person}{Rafa Mantiuk}, \bibinfo{person}{Kil~Joong
  Kim}, \bibinfo{person}{Allan~G. Rempel}, {and} \bibinfo{person}{Wolfgang
  Heidrich}.} \bibinfo{year}{2011}\natexlab{a}.
\newblock \showarticletitle{HDR-VDP-2: A Calibrated Visual Metric for
  Visibility and Quality Predictions in All Luminance Conditions}.
\newblock \bibinfo{journal}{\emph{ACM Trans. Graph.}} \bibinfo{volume}{30},
  \bibinfo{number}{4}, Article \bibinfo{articleno}{40} (\bibinfo{date}{July}
  \bibinfo{year}{2011}), \bibinfo{numpages}{14}~pages.
\newblock
\showISSN{0730-0301}
\urldef\tempurl%
\url{https://doi.org/10.1145/2010324.1964935}
\showDOI{\tempurl}


\bibitem[\protect\citeauthoryear{Mantiuk, Kim, Rempel, and Heidrich}{Mantiuk
  et~al\mbox{.}}{2011b}]%
        {hdr_vdp2}
\bibfield{author}{\bibinfo{person}{Rafa\l{} Mantiuk},
  \bibinfo{person}{Kil~Joong Kim}, \bibinfo{person}{Allan~G. Rempel}, {and}
  \bibinfo{person}{Wolfgang Heidrich}.} \bibinfo{year}{2011}\natexlab{b}.
\newblock \showarticletitle{HDR-VDP-2: A Calibrated Visual Metric for
  Visibility and Quality Predictions in All Luminance Conditions}.
\newblock  \bibinfo{volume}{30}, \bibinfo{number}{4}, Article
  \bibinfo{articleno}{40} (\bibinfo{date}{July} \bibinfo{year}{2011}),
  \bibinfo{numpages}{14}~pages.
\newblock
\showISSN{0730-0301}
\urldef\tempurl%
\url{https://doi.org/10.1145/2010324.1964935}
\showDOI{\tempurl}


\bibitem[\protect\citeauthoryear{Marnerides, Bashford-Rogers, Hatchett, and
  Debattista}{Marnerides et~al\mbox{.}}{2018}]%
        {marnerides2018expandnet}
\bibfield{author}{\bibinfo{person}{Demetris Marnerides},
  \bibinfo{person}{Thomas Bashford-Rogers}, \bibinfo{person}{Jonathan
  Hatchett}, {and} \bibinfo{person}{Kurt Debattista}.}
  \bibinfo{year}{2018}\natexlab{}.
\newblock \showarticletitle{ExpandNet: A deep convolutional neural network for
  high dynamic range expansion from low dynamic range content}. In
  \bibinfo{booktitle}{\emph{Computer Graphics Forum}},
  Vol.~\bibinfo{volume}{37}. Wiley Online Library, \bibinfo{pages}{37--49}.
\newblock


\bibitem[\protect\citeauthoryear{Maurer and Bruhn}{Maurer and Bruhn}{2018}]%
        {maurer2018proflow}
\bibfield{author}{\bibinfo{person}{Daniel Maurer} {and}
  \bibinfo{person}{Andrés Bruhn}.} \bibinfo{year}{2018}\natexlab{}.
\newblock \bibinfo{title}{ProFlow: Learning to Predict Optical Flow}.
\newblock
\newblock
\showeprint[arxiv]{1806.00800}~[cs.CV]


\bibitem[\protect\citeauthoryear{Mobahi, Collobert, and Weston}{Mobahi
  et~al\mbox{.}}{2009}]%
        {temp}
\bibfield{author}{\bibinfo{person}{Hossein Mobahi}, \bibinfo{person}{Ronan
  Collobert}, {and} \bibinfo{person}{Jason Weston}.}
  \bibinfo{year}{2009}\natexlab{}.
\newblock \showarticletitle{Deep Learning from Temporal Coherence in Video}. In
  \bibinfo{booktitle}{\emph{Proceedings of the 26th Annual International
  Conference on Machine Learning}} (Montreal, Quebec, Canada)
  \emph{(\bibinfo{series}{ICML '09})}. \bibinfo{publisher}{Association for
  Computing Machinery}, \bibinfo{address}{New York, NY, USA},
  \bibinfo{pages}{737–744}.
\newblock
\showISBNx{9781605585161}
\urldef\tempurl%
\url{https://doi.org/10.1145/1553374.1553469}
\showDOI{\tempurl}


\bibitem[\protect\citeauthoryear{Nayar and Mitsunaga}{Nayar and
  Mitsunaga}{2000}]%
        {nayar2000high}
\bibfield{author}{\bibinfo{person}{Shree~K Nayar} {and} \bibinfo{person}{Tomoo
  Mitsunaga}.} \bibinfo{year}{2000}\natexlab{}.
\newblock \showarticletitle{High dynamic range imaging: Spatially varying pixel
  exposures}. In \bibinfo{booktitle}{\emph{Proceedings IEEE Conference on
  Computer Vision and Pattern Recognition. CVPR 2000 (Cat. No. PR00662)}},
  Vol.~\bibinfo{volume}{1}. IEEE, \bibinfo{pages}{472--479}.
\newblock


\bibitem[\protect\citeauthoryear{Oh, Lee, Tai, and Kweon}{Oh
  et~al\mbox{.}}{2014}]%
        {oh2014robust}
\bibfield{author}{\bibinfo{person}{Tae-Hyun Oh}, \bibinfo{person}{Joon-Young
  Lee}, \bibinfo{person}{Yu-Wing Tai}, {and} \bibinfo{person}{In~So Kweon}.}
  \bibinfo{year}{2014}\natexlab{}.
\newblock \showarticletitle{Robust high dynamic range imaging by rank
  minimization}.
\newblock \bibinfo{journal}{\emph{IEEE transactions on pattern analysis and
  machine intelligence}} \bibinfo{volume}{37}, \bibinfo{number}{6}
  (\bibinfo{year}{2014}), \bibinfo{pages}{1219--1232}.
\newblock


\bibitem[\protect\citeauthoryear{{Ranjan} and {Black}}{{Ranjan} and
  {Black}}{2017}]%
        {pyramid_flow}
\bibfield{author}{\bibinfo{person}{A. {Ranjan}} {and} \bibinfo{person}{M.~J.
  {Black}}.} \bibinfo{year}{2017}\natexlab{}.
\newblock \showarticletitle{Optical Flow Estimation Using a Spatial Pyramid
  Network}. In \bibinfo{booktitle}{\emph{2017 IEEE Conference on Computer
  Vision and Pattern Recognition (CVPR)}}. \bibinfo{pages}{2720--2729}.
\newblock


\bibitem[\protect\citeauthoryear{Ronneberger, Fischer, and Brox}{Ronneberger
  et~al\mbox{.}}{2015}]%
        {ronneberger2015unet}
\bibfield{author}{\bibinfo{person}{Olaf Ronneberger}, \bibinfo{person}{Philipp
  Fischer}, {and} \bibinfo{person}{Thomas Brox}.}
  \bibinfo{year}{2015}\natexlab{}.
\newblock \bibinfo{title}{U-Net: Convolutional Networks for Biomedical Image
  Segmentation}.
\newblock
\newblock
\showeprint[arxiv]{1505.04597}~[cs.CV]


\bibitem[\protect\citeauthoryear{Sanakoyeu, Kotovenko, Lang, and
  Ommer}{Sanakoyeu et~al\mbox{.}}{2018}]%
        {sanakoyeu2018styleaware}
\bibfield{author}{\bibinfo{person}{Artsiom Sanakoyeu}, \bibinfo{person}{Dmytro
  Kotovenko}, \bibinfo{person}{Sabine Lang}, {and} \bibinfo{person}{Bjorn
  Ommer}.} \bibinfo{year}{2018}\natexlab{}.
\newblock \showarticletitle{A Style-Aware Content Loss for Real-time HD Style
  Transfer}. In \bibinfo{booktitle}{\emph{Proceedings of the European
  Conference on Computer Vision (ECCV)}}.
\newblock


\bibitem[\protect\citeauthoryear{Seger, Apel, and H{\"o}fflinger}{Seger
  et~al\mbox{.}}{1999}]%
        {seger1999hdrc}
\bibfield{author}{\bibinfo{person}{Ulrich Seger}, \bibinfo{person}{Uwe Apel},
  {and} \bibinfo{person}{Bernd H{\"o}fflinger}.}
  \bibinfo{year}{1999}\natexlab{}.
\newblock \showarticletitle{HDRC-imagers for natural visual perception}.
\newblock \bibinfo{journal}{\emph{Handbook of Computer Vision and Application}}
   \bibinfo{volume}{1} (\bibinfo{year}{1999}), \bibinfo{pages}{223--235}.
\newblock


\bibitem[\protect\citeauthoryear{Sen, Kalantari, Yaesoubi, Darabi, Goldman, and
  Shechtman}{Sen et~al\mbox{.}}{2012}]%
        {sen2012robust}
\bibfield{author}{\bibinfo{person}{Pradeep Sen}, \bibinfo{person}{Nima~Khademi
  Kalantari}, \bibinfo{person}{Maziar Yaesoubi}, \bibinfo{person}{Soheil
  Darabi}, \bibinfo{person}{Dan~B Goldman}, {and} \bibinfo{person}{Eli
  Shechtman}.} \bibinfo{year}{2012}\natexlab{}.
\newblock \showarticletitle{Robust patch-based hdr reconstruction of dynamic
  scenes.}
\newblock \bibinfo{journal}{\emph{ACM Trans. Graph.}} \bibinfo{volume}{31},
  \bibinfo{number}{6} (\bibinfo{year}{2012}), \bibinfo{pages}{203--1}.
\newblock


\bibitem[\protect\citeauthoryear{Simonyan and Zisserman}{Simonyan and
  Zisserman}{2014}]%
        {karen}
\bibfield{author}{\bibinfo{person}{Karen Simonyan} {and}
  \bibinfo{person}{Andrew Zisserman}.} \bibinfo{year}{2014}\natexlab{}.
\newblock \showarticletitle{Two-Stream Convolutional Networks for Action
  Recognition in Videos}. In \bibinfo{booktitle}{\emph{Proceedings of the 27th
  International Conference on Neural Information Processing Systems - Volume
  1}} (Montreal, Canada) \emph{(\bibinfo{series}{NIPS'14})}.
  \bibinfo{publisher}{MIT Press}, \bibinfo{address}{Cambridge, MA, USA},
  \bibinfo{pages}{568–576}.
\newblock


\bibitem[\protect\citeauthoryear{Thasarathan, Nazeri, and Ebrahimi}{Thasarathan
  et~al\mbox{.}}{2019}]%
        {thasarathan2019automatic}
\bibfield{author}{\bibinfo{person}{Harrish Thasarathan},
  \bibinfo{person}{Kamyar Nazeri}, {and} \bibinfo{person}{Mehran Ebrahimi}.}
  \bibinfo{year}{2019}\natexlab{}.
\newblock \showarticletitle{Automatic temporally coherent video colorization}.
  In \bibinfo{booktitle}{\emph{2019 16th Conference on Computer and Robot
  Vision (CRV)}}. IEEE, \bibinfo{pages}{189--194}.
\newblock


\bibitem[\protect\citeauthoryear{Tocci, Kiser, Tocci, and Sen}{Tocci
  et~al\mbox{.}}{[n.d.]}]%
        {tocciversatile}
\bibfield{author}{\bibinfo{person}{MD Tocci}, \bibinfo{person}{C Kiser},
  \bibinfo{person}{N Tocci}, {and} \bibinfo{person}{P Sen}.}
  \bibinfo{year}{[n.d.]}\natexlab{}.
\newblock \bibinfo{title}{A Versatile HDR Video Production System. ACM TOG 30
  (4), 41: 1--41: 10 (Jul 2011)}.
\newblock
\newblock


\bibitem[\protect\citeauthoryear{Tocci, Kiser, Tocci, and Sen}{Tocci
  et~al\mbox{.}}{2011}]%
        {tocci2011versatile}
\bibfield{author}{\bibinfo{person}{Michael~D Tocci}, \bibinfo{person}{Chris
  Kiser}, \bibinfo{person}{Nora Tocci}, {and} \bibinfo{person}{Pradeep Sen}.}
  \bibinfo{year}{2011}\natexlab{}.
\newblock \showarticletitle{A versatile HDR video production system}.
\newblock \bibinfo{journal}{\emph{ACM Transactions on Graphics (TOG)}}
  \bibinfo{volume}{30}, \bibinfo{number}{4} (\bibinfo{year}{2011}),
  \bibinfo{pages}{1--10}.
\newblock


\bibitem[\protect\citeauthoryear{Tomaszewska and Mantiuk}{Tomaszewska and
  Mantiuk}{2007}]%
        {tomaszewska2007image}
\bibfield{author}{\bibinfo{person}{Anna Tomaszewska} {and}
  \bibinfo{person}{Radoslaw Mantiuk}.} \bibinfo{year}{2007}\natexlab{}.
\newblock \showarticletitle{Image registration for multi-exposure high dynamic
  range image acquisition}.
\newblock  (\bibinfo{year}{2007}).
\newblock


\bibitem[\protect\citeauthoryear{Ward}{Ward}{2003}]%
        {ward2003fast}
\bibfield{author}{\bibinfo{person}{Greg Ward}.}
  \bibinfo{year}{2003}\natexlab{}.
\newblock \showarticletitle{Fast, robust image registration for compositing
  high dynamic range photographs from hand-held exposures}.
\newblock \bibinfo{journal}{\emph{Journal of graphics tools}}
  \bibinfo{volume}{8}, \bibinfo{number}{2} (\bibinfo{year}{2003}),
  \bibinfo{pages}{17--30}.
\newblock


\bibitem[\protect\citeauthoryear{Xu, Huang, Cheng, Liu, Zhu, Xu, and Shao}{Xu
  et~al\mbox{.}}{2019}]%
        {xu2019noisyasclean}
\bibfield{author}{\bibinfo{person}{Jun Xu}, \bibinfo{person}{Yuan Huang},
  \bibinfo{person}{Ming-Ming Cheng}, \bibinfo{person}{Li Liu},
  \bibinfo{person}{Fan Zhu}, \bibinfo{person}{Zhou Xu}, {and}
  \bibinfo{person}{Ling Shao}.} \bibinfo{year}{2019}\natexlab{}.
\newblock \bibinfo{title}{Noisy-As-Clean: Learning Self-supervised Denoising
  from the Corrupted Image}.
\newblock
\newblock
\showeprint[arxiv]{1906.06878}~[cs.CV]


\bibitem[\protect\citeauthoryear{Zhang and Lalonde}{Zhang and Lalonde}{2017}]%
        {zhang-iccv-17}
\bibfield{author}{\bibinfo{person}{Jinsong Zhang} {and}
  \bibinfo{person}{Jean-Fran\c{c}ois Lalonde}.}
  \bibinfo{year}{2017}\natexlab{}.
\newblock \showarticletitle{Learning High Dynamic Range from Outdoor
  Panoramas}. In \bibinfo{booktitle}{\emph{IEEE International Conference on
  Computer Vision}}.
\newblock


\bibitem[\protect\citeauthoryear{Zhang, Deshpande, and Chen}{Zhang
  et~al\mbox{.}}{2010}]%
        {zhang2010denoising}
\bibfield{author}{\bibinfo{person}{Li Zhang}, \bibinfo{person}{Alok Deshpande},
  {and} \bibinfo{person}{Xin Chen}.} \bibinfo{year}{2010}\natexlab{}.
\newblock \showarticletitle{Denoising vs. deblurring: HDR imaging techniques
  using moving cameras}. In \bibinfo{booktitle}{\emph{2010 IEEE Computer
  Society Conference on Computer Vision and Pattern Recognition}}. IEEE,
  \bibinfo{pages}{522--529}.
\newblock


\bibitem[\protect\citeauthoryear{Zhang, Isola, and Efros}{Zhang
  et~al\mbox{.}}{2016}]%
        {color}
\bibfield{author}{\bibinfo{person}{Richard Zhang}, \bibinfo{person}{Phillip
  Isola}, {and} \bibinfo{person}{Alexei~A. Efros}.}
  \bibinfo{year}{2016}\natexlab{}.
\newblock \showarticletitle{Colorful Image Colorization}.
\newblock \bibinfo{journal}{\emph{CoRR}}  \bibinfo{volume}{abs/1603.08511}
  (\bibinfo{year}{2016}).
\newblock
\showeprint[arxiv]{1603.08511}
\urldef\tempurl%
\url{http://arxiv.org/abs/1603.08511}
\showURL{%
\tempurl}


\bibitem[\protect\citeauthoryear{Zhao, Shi, Fernandez-Cull, Yeung, and
  Raskar}{Zhao et~al\mbox{.}}{2015}]%
        {zhao2015unbounded}
\bibfield{author}{\bibinfo{person}{Hang Zhao}, \bibinfo{person}{Boxin Shi},
  \bibinfo{person}{Christy Fernandez-Cull}, \bibinfo{person}{Sai-Kit Yeung},
  {and} \bibinfo{person}{Ramesh Raskar}.} \bibinfo{year}{2015}\natexlab{}.
\newblock \showarticletitle{Unbounded high dynamic range photography using a
  modulo camera}. In \bibinfo{booktitle}{\emph{2015 IEEE International
  Conference on Computational Photography (ICCP)}}. IEEE,
  \bibinfo{pages}{1--10}.
\newblock


\bibitem[\protect\citeauthoryear{Zheng, Li, Zhu, Wu, and Rahardja}{Zheng
  et~al\mbox{.}}{2013}]%
        {zheng2013hybrid}
\bibfield{author}{\bibinfo{person}{Jinghong Zheng}, \bibinfo{person}{Zhengguo
  Li}, \bibinfo{person}{Zijian Zhu}, \bibinfo{person}{Shiqian Wu}, {and}
  \bibinfo{person}{Susanto Rahardja}.} \bibinfo{year}{2013}\natexlab{}.
\newblock \showarticletitle{Hybrid patching for a sequence of differently
  exposed images with moving objects}.
\newblock \bibinfo{journal}{\emph{IEEE Transactions on Image Processing}}
  \bibinfo{volume}{22}, \bibinfo{number}{12} (\bibinfo{year}{2013}),
  \bibinfo{pages}{5190--5201}.
\newblock


\bibitem[\protect\citeauthoryear{Zhou, Liang, Song, Yu, Wang, Zhang, Yu, and
  Zhang}{Zhou et~al\mbox{.}}{2019}]%
        {lipschitz}
\bibfield{author}{\bibinfo{person}{Zhiming Zhou}, \bibinfo{person}{Jiadong
  Liang}, \bibinfo{person}{Yuxuan Song}, \bibinfo{person}{Lantao Yu},
  \bibinfo{person}{Hongwei Wang}, \bibinfo{person}{Weinan Zhang},
  \bibinfo{person}{Yong Yu}, {and} \bibinfo{person}{Zhihua Zhang}.}
  \bibinfo{year}{2019}\natexlab{}.
\newblock \showarticletitle{Lipschitz Generative Adversarial Nets}.
\newblock \bibinfo{journal}{\emph{CoRR}}  \bibinfo{volume}{abs/1902.05687}
  (\bibinfo{year}{2019}).
\newblock
\showeprint[arxiv]{1902.05687}
\urldef\tempurl%
\url{http://arxiv.org/abs/1902.05687}
\showURL{%
\tempurl}


\bibitem[\protect\citeauthoryear{{Zhou Wang}, {Bovik}, {Sheikh}, and
  {Simoncelli}}{{Zhou Wang} et~al\mbox{.}}{2004}]%
        {ssim}
\bibfield{author}{\bibinfo{person}{{Zhou Wang}}, \bibinfo{person}{A.~C.
  {Bovik}}, \bibinfo{person}{H.~R. {Sheikh}}, {and} \bibinfo{person}{E.~P.
  {Simoncelli}}.} \bibinfo{year}{2004}\natexlab{}.
\newblock \showarticletitle{Image quality assessment: from error visibility to
  structural similarity}.
\newblock \bibinfo{journal}{\emph{IEEE Transactions on Image Processing}}
  \bibinfo{volume}{13}, \bibinfo{number}{4} (\bibinfo{year}{2004}),
  \bibinfo{pages}{600--612}.
\newblock
\urldef\tempurl%
\url{https://doi.org/10.1109/TIP.2003.819861}
\showDOI{\tempurl}


\end{thebibliography}

\appendix

\end{document}